\documentclass[iop]{emulateapj}

\usepackage{times,color,natbib,url,amsmath}
\usepackage{graphicx,float,psfrag}
\usepackage{tabularx}
\usepackage{latexsym,amsmath,amssymb}
\usepackage{natbib}
\usepackage{verbatim}
\usepackage{multirow}

\bibpunct{(}{)}{;}{a}{}{,}

\newcommand {\xmm} {\textsl{XMM-Newton}}

\newcommand {\swift} {\textsl{Swift}}
\newcommand {\rxte} {\textsl{RXTE}}
\newcommand {\fermi} {\textsl{Fermi}}

\def \rsun {\ifmmode$R$_{\odot}\else R$_{\odot}$}

\def \hcm {\hbox {\ifmmode $ atoms cm$^{-2}\else atoms cm$^{-2}$\fi}}

\def\approxgt{\mathrel{\hbox{\rlap{\lower.55ex \hbox {$\sim$}}
        \kern-.3em \raise.4ex \hbox{$>$}}}}
\def\approxlt{\mathrel{\hbox{\rlap{\lower.55ex \hbox {$\sim$}}
        \kern-.3em \raise.4ex \hbox{$<$}}}}

\newcommand {\degree} {$^{\circ}$}

\newcommand {\epeak} {$E_{\rm peak}$}

\def \src {SGR J1550-5418}

\begin{document}

\title{Time resolved spectroscopy of \src\ bursts detected with Fermi/GBM}

%
%
\author{G.~Younes$^{1,2}$, C.~Kouveliotou$^{3,2}$,
  A.~J.~van~der~Horst$^{4}$, M.~G.~Baring$^{5}$, J.~Granot$^{6}$,
  A.~L.~Watts$^{4}$, P.~N.~Bhat$^{7}$, A.~Collazzi$^{3,8}$,
  N.~Gehrels$^9$, N.~Gorgone$^{7}$, E.~G\"o\u{g}\"u\c{s}$^{10}$,
  D.~Gruber$^{11}$, S.~Grunblatt$^{12}$, D.~Huppenkothen$^4$,
  Y.~Kaneko$^{10}$, A.~von~Kienlin$^{11}$, M.~van der Klis$^{4}$,
  L.~Lin$^{10}$, J.~Mcenery$^{9}$, T.~van Putten$^{4}$,
  R.~A.~M.~J.~Wijers$^4$}

\affil{
 $^1$ Universities Space Research Association, 6767 Old Madison Pike, Suite 450, Huntsville, AL 35806, USA \\
 $^2$ NSSTC, 320 Sparkman Drive, Huntsville, AL 35805, USA \\
 $^3$ Astrophysics Office, ZP 12, NASA-Marshall Space Flight Center, Huntsville, AL 35812, USA \\
 $^4$ Astronomical Institute "Anton Pannekoek," University of Amsterdam, Postbus 94249, 1090 GE Amsterdam, The Netherlands \\
 $^5$ Department of Physics and Astronomy, Rice University, MS-108, P.O. Box 1892, Houston, TX 77251, USA \\
 $^6$ Department of Natural Sciences, The Open University of Israel, 1 University Road, P.O. Box 808, Ra\'anana 43537, Israel \\
 $^7$ University of Alabama in Huntsville CSPAR, 320 Sparkman Dr. Huntsville AL, 35805, USA \\
 $^{8}$ NASA Postdoctoral Program, USA \\
 $^{9}$ NASA Goddard Space Flight Center, Greenbelt, MD 20771, USA \\
 $^{10}$ Sabanc\i~University, Orhanl\i-Tuzla, \.Istanbul 34956, Turkey \\
 $^{11}$ Max Planck Institute for Extraterrestrial Physics, Giessenbachstrasse, Postfach 1312, 85748 Garching, Germany \\
 $^{12}$ University of Hawaii at Manoa, 2500 Campus Rd, Honolulu, HI 96822, USA \\
}

\date{}
%
%

\begin{abstract}

We report on time-resolved spectroscopy of the 63 brightest bursts of
\src, detected with {\it Fermi}/Gamma-ray Burst Monitor during its
2008-2009 intense bursting episode. We performed spectral analysis
down to 4~ms time-scales, to characterize the spectral evolution of
the bursts. Using a Comptonized model, we find that the peak energy,
\epeak, anti-correlates with flux, while the low-energy photon index
remains constant at $\sim-0.8$ up to a flux limit
$F\approx10^{-5}$~erg~s$^{-1}$~cm$^{-2}$. Above this flux value the
\epeak$-$flux correlation changes sign, and the index positively
correlates with flux reaching $\sim$1 at the highest fluxes. Using a
two black-body model, we find that the areas and fluxes of the two
emitting regions correlate positively. Further, we study here for the
first time, the evolution of the temperatures and areas as a function
of flux. We find that the area$-kT$ relation follows lines of constant
luminosity at the lowest fluxes, $R^2\propto kT^{-4}$, with a break at
higher fluxes ($F>10^{-5.5}$~erg~s$^{-1}$~cm$^{-2}$). The area of the
high$-kT$ component increases with flux while its temperature
decreases, which we interpret as due to an adiabatic cooling
process. The area of the low$-kT$ component, on the other hand,
appears to saturate at the highest fluxes, towards $R_{\rm
  max}\approx30$~km. Assuming that crust quakes are responsible for
SGR bursts and considering $R_{\rm max}$ as the maximum radius of the 
emitting photon-pair plasma fireball, we relate this saturation radius
to a minimum excitation radius of the magnetosphere, and put a lower
limit on the internal magnetic field of \src, $B_{\rm
  int}\gtrsim4.5\times10^{15}$~G.

\end{abstract}

\section{Introduction}
\label{Sec:Intro}

Soft gamma repeaters (SGRs) represent a small subset of the isolated
neutron star (NS) population. They are characterized by short
($\sim$0.1~s), bright ($10^{38}-10^{41}$~erg) bursts of hard
X-ray/soft gamma-ray emission. Very rarely, they emit giant flares 
with extreme energies ($10^{44}-10^{46}$~erg), characterized
by an initial very short, hard spike and a decaying tail lasting
several minutes. Intermediate duration and luminosity bursts were also
recorded for a few SGRs \citep[e.g.,][]{kouveliotou01ApJ:IF1900}. Most
SGRs are bright X-ray sources, with luminosities significantly larger
than those expected from rotational energy losses. Their spin periods
are clustered between 2-12~s, and their large spin-down rates imply
very high surface dipole magnetic fields of $10^{14}-10^{15}$~G
\citep[e.g., ][but also see
\citealt{rea13:0418}]{kouv98Natur:1806,kouveliotou99ApJ:1900}. The
above properties argue strongly in favor of the nature of SGRs as very
strongly magnetized neutron stars, i.e., magnetars
\citep{duncan92ApJmagnetars, paczynski92AcA:magnetars}. 

Another class of isolated NS, the Anomalous X-ray Pulsars (AXPs), show
very similar characteristics to SGRs, including spin down rates
leading to extreme magnetic fields. AXPs were discovered as bright
persistent X-ray sources \citep{mereghetti95ApJ:AXPs}, however, 
\citet{gavriil02Natur:AXPs} reported  in 2002 the discovery of
SGR-like X-ray bursts from AXP 1E~1048.1$-$5937. Many other burst
detections from AXPs followed (with the exception of giant flares),
implying a possible evolutionary link between these two classes of
magnetars \citep[see][for reviews]{woods05ApJ:xte1810,
  mereghetti95ApJ:AXPs, perna11ApJ}.

Magnetars become burst active randomly with active periods lasting for
relatively small time-intervals of weeks to months, between several
years of quiescence. The total energy released during these episodes
varies from source to source, and between active episodes of the same 
source \citep[e.g., ][]{kouveliotou93Natur:1900}. Past spectral
analyses of SGR short and intermediate bursts revealed that an
optically thin thermal bremsstrahlung (OTTB) model fits the data above
$\sim15$~keV well, with temperatures in the range of 20-40~keV
\citep{aptekar01ApJS:SGR}. However, the OTTB model overestimates
fluxes below 15 keV \citep[e.g., ][]{fenimore94ApJ:1806,
  feroci04ApJ:1900}. Models consisting of a two black-body
(2BB) or a power-law (PL) with a high-energy cutoff (Comptonized
model; $f=A\exp{[-E(2+\lambda)/E_{\rm peak}](E/E_{\rm
      piv})^\lambda}$, where $E_{\rm pivot}$ is set to 30~keV) were
shown to best fit the 1-150~keV broad-band spectra of SGR bursts
\citep[e.g., ][V12 hereafter]{feroci04ApJ:1900,olive04ApJ:1900,
  lin12ApJ:1550,vanderhorst12ApJ:1550}.

The origin and radiative mechanism of the persistent and burst
emission from magnetars are not yet fully understood. In the magnetar
framework, the decay of the magnetic field powers the quiescent
emission through fracturing of the NS crust and sub-surface heating 
\citep{thompson96ApJ:magnetar}. Bursts occur when a build-up of
magnetic stresses on the NS crust from a twisted magnetic field causes
the crust to crack, ejecting hot plasma into the magnetosphere
\citep{thompson95MNRAS:GF,thompson02ApJ:magnetars}. Alternatively,
magnetar bursts could result from magnetic field reconnection in the
magnetosphere \citep[][see also \citealt{gill10MNRAS:SGRGF}]{
  lyutikov03MNRAS:recon}. Although the distinction between the two
models is extremely challenging to measure, detailed statistical,
temporal, and spectral studies of magnetar bursts have been
effectively used for model comparisons \citep[e.g.,][V12]{
  gogus99apjl:1900, gogus00ApJ:1806, woods05ApJ:xte1810,
  israel08ApJ:1900, lin12ApJ:1550, lin11ApJ:0501,
  savchenko10AA:1547}.
  
\src, which is the subject of this study, was suggested as a magnetar
candidate in the supernova remnant G327.24$-$0.13 based on its X-ray
spectral shape and varying flux \citep{gelfand07ApJ:1547}. Subsequent
radio observations revealed its magnetar nature with the discovery of
radio pulsations at a period of 2.1~s, and
$\dot{P}=2.3\times10^{-11}$~s$^{-1}$, implying a surface field
strength of $2.2\times10^{14}$~G \citep{camilo07ApJ:1547}. A \xmm\
observation caught the source in outburst, which allowed the detection
of X-ray pulsations at the same spin period \citep{
  halpern08ApJ:1547}. No bursts from \src\ were detected until 2008
when the source entered three active episodes: October 2008
\citep{israel10MNRAS:1547, vonkienlin12ApJ:1550}, January 2009, and 
March 2009. The most prolific activity took place on 22 January 2009,
when the source emitted hundreds of bursts in 24 hours. These latter
bursts were detected by multiple high-energy instruments such as the
\fermi/GBM, \swift/BAT, \rxte/PCA, and {\sl INTEGRAL}, and their
temporal and spectral properties have been studied extensively
\citep[V12;][]{mereghetti09ApJ:1547,savchenko10AA:1547,kaneko10ApJ:1550,scholz11ApJ:1547}.

Here we report on the time-resolved spectroscopy of the brightest
bursts from \src\ seen with \fermi/GBM, which is ideal for such
analyses owing to its high time-resolution and spectral
capabilities. In particular, we concentrate on the spectral evolution
of these bursts using the two spectral models established earlier as
best for the source (V12), i.e., Comptonized model and the 2BB
model. In Section~\ref{datSel}, we describe the data selection and
burst identification technique. We present our results on the
time-resolved spectroscopy in Section~\ref{timeResSpec}. In
Section~\ref{disc}, we first compare our results to the ones derived
with time-integrated spectroscopy, and then to previous results on
time-resolved spectroscopy of different sources. We interpret our
results in the context of the current theoretical models for the
origin of magnetar bursts, and the radiative processes that accompany
them. Finally, we conclude with a few remarks in
Section~\ref{conclusion}.

\section{Data selection}
\label{datSel}

The Gamma-ray Burst Monitor (GBM) onboard {\sl Fermi} telescope has a
continuous broadband energy coverage (8~keV$-$40~MeV) of the earth
un-occulted sky. It consists of 12 NaI detectors (8$-$1000~keV) and 2
Bismuth Germanate (BGO) detectors (0.2$-$40~MeV). We analyzed bursts with
time-tagged photon event (TTE) data (2~$\mu$s and 128 energy
channels), which at the time of the outburst, were provided 30~s
before a trigger occurs, to 300~s post-trigger. See
\citet{meegan09ApJ:gbm} for a detailed description of the instrument
and data products.

For our entire analysis, we used the NaI detectors with an angle to
the source smaller than 50\degree\ to avoid attenuation effects. We
also excluded any detectors blocked by the Fermi Large Area Telescope
or by the spacecraft radiators or solar panels. The BGO detectors were
not used as there was no obvious emission in the NaIs above 200 keV
(see V12 for details).

The time-integrated spectroscopy of \src\ bursts detected with GBM has
been presented in V12, \citet{vonkienlin12ApJ:1550} and Collazzi et
al. (in preparation). To ensure that the bursts studied here have enough
statistics for time-resolved spectroscopy we selected bursts with a
fluence $>10^{-6}$~erg~cm$^{-2}$ or with an average flux
$>5\times10^{-6}$~erg~s$^{-1}$~cm$^{-2}$ (both in the 8$-$200 keV
range), resulting in an initial sample of 63 bursts (the log of these
bursts will be presented in the Collazzi et al. 2013, in prep.).

\begin{figure}[t]
\begin{center}
\includegraphics[angle=0,width=0.49\textwidth]{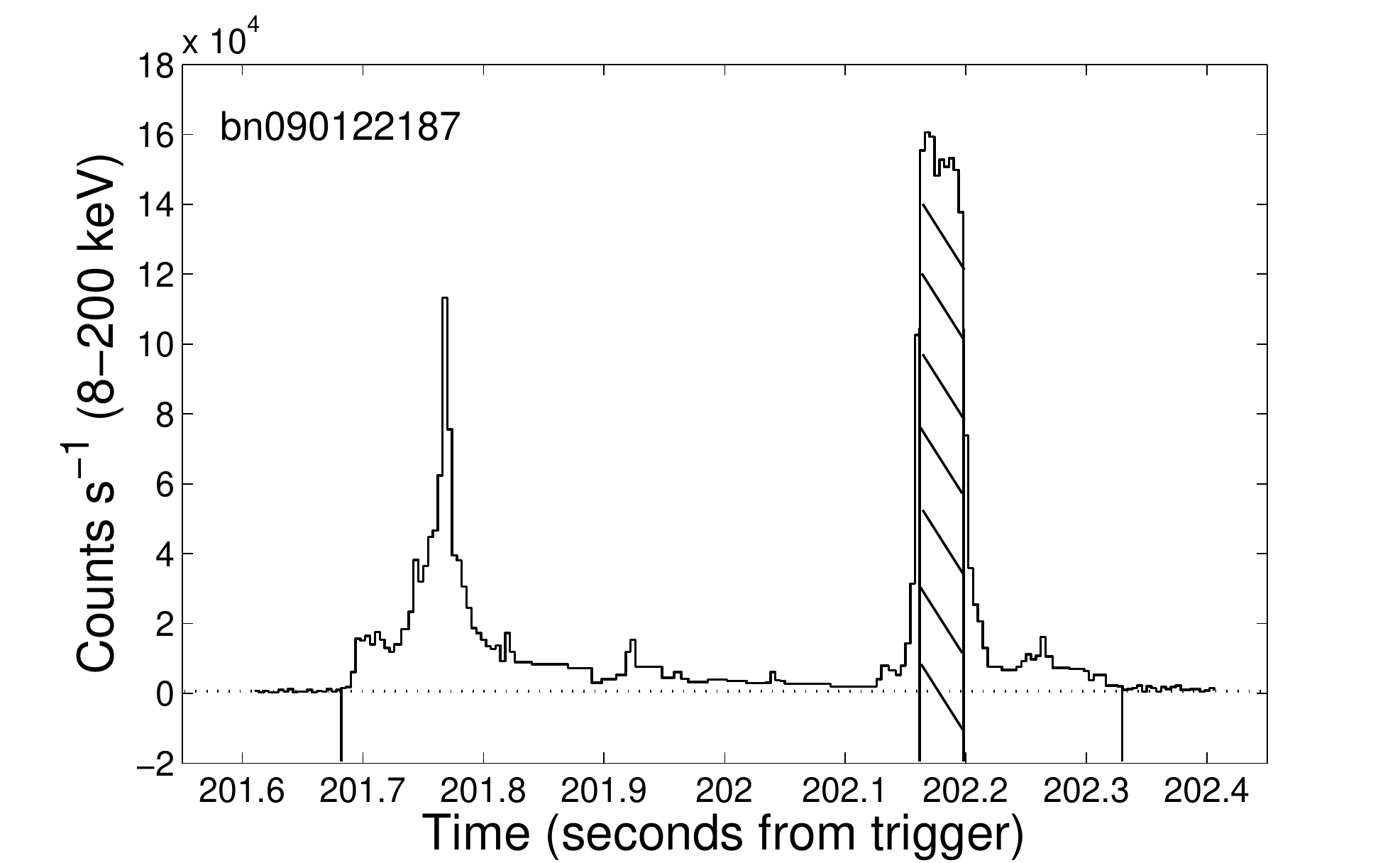}\\
\includegraphics[angle=0,width=0.49\textwidth]{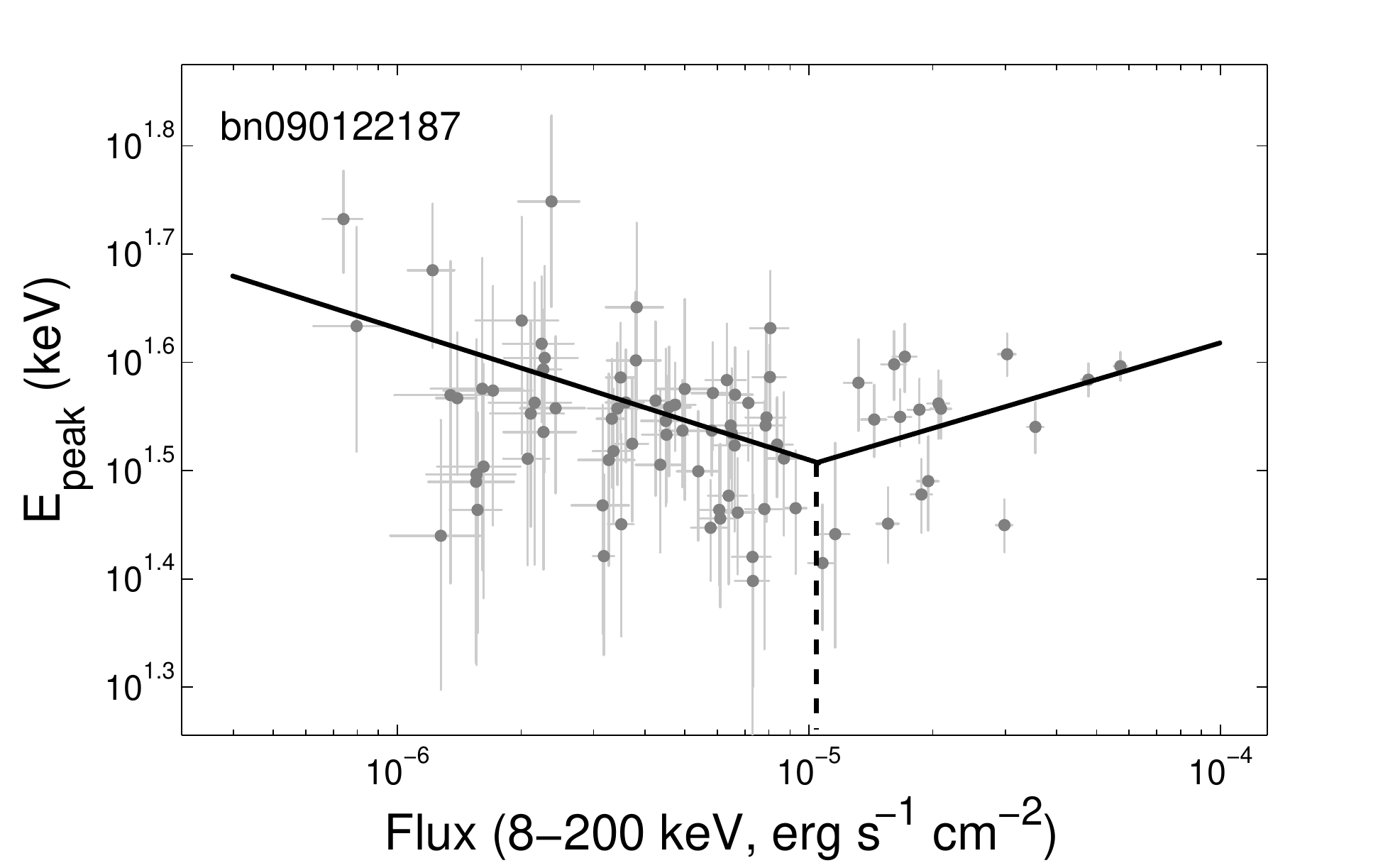}
\caption{{\sl Upper panel.} Example of a {\it Fermi}-GBM bright burst
  (untriggered) from \src. Each data bin is fit with either a
  COMPT or a 2BB model. The shaded region represents saturated
  intervals, which were not included in our analysis. The dotted 
  line represents the background. {\sl Lower panel.} COMPT \epeak\ as
  a function of flux for the same burst. The solid line is the best
  fit BPL model. The dashed line indicates where the break occurs in
  flux space, i.e., $F\approx10^{-5}$~erg~s$^{-1}$~cm$^{-2}$. The two
  PL indices are $-0.12$ and $0.11$, for the parts below and above the
  break, respectively.}
\label{epkExa}
\end{center}
\end{figure}

The time intervals used for time-integrated spectroscopy of \src\
bursts in the aforementioned papers were chosen to be close to T$_{90}$
\citep[the time interval in which the central 90\% of the burst
counts are accumulated;][]{kouveliotou93ApJ:GRBs}. While this is adequate
for such analysis, it is not ideal when performing time-resolved
spectroscopy. In the time-integrated analysis the spectrum is
dominated by the brightest time intervals, while the low level
emission intervals do not contribute significantly to the total burst
spectrum. These intervals, however, can contain important information
for studying the burst spectral evolution, so we proceeded to identify
the faintest, statistically significant intervals that could be used
for spectral analysis for each burst. First we inspected the light
curves plotted with 4\,ms bins of all selected bursts using the GBM
detector with the smallest angle to the source (brightest
detector). Then we searched for the start and end times of a given
burst by looking at intervals of 0.1~s before and after the start and
end times of T$_{90}$ for each burst.  We required that the count
rates remained at least $2\sigma$ continuously above the background,
to ensure that we included the faint wings of each event. The
resulting bins were used for spectral analysis, either in 4\,ms
resolution or binned with a coarser resolution, to achieve $>3\sigma$
significance (see also section~\ref{compMod}). We note that saturation
intervals, e.g., Figure~\ref{epkExa}, were excluded from all analyses
(V12).

\section{Time-resolved spectroscopy}
\label{timeResSpec}

The time-integrated spectroscopy of \src\ bursts \citep[][V12,
Collazzi et al. in prep]{vonkienlin12ApJ:1550} has shown that their
spectra are well fit by a Comptonized model (COMPT). The other model
to best fit the data is a combination of two blackbody (BB) functions
(2BB model), except for the bursts in October 2008. During that time
frame, however, there were no bursts that met our fluence and flux
selection criteria, therefore we were able to use COMPT and 2BB models
in our time-resolved spectroscopy. In this section we discuss the
analysis with both spectral models separately. For the spectral
analysis we have used the software package {\it RMFIT (v4.3)}, and we
generated the detector response matrices with {\it GBMRSP
  v1.81}. Given the low number of counts per time bin we have
minimized the Castor C-statistic to obtain the best spectral fit.

\begin{figure*}[th!]
\begin{center}
\includegraphics[angle=0,width=0.48\textwidth]{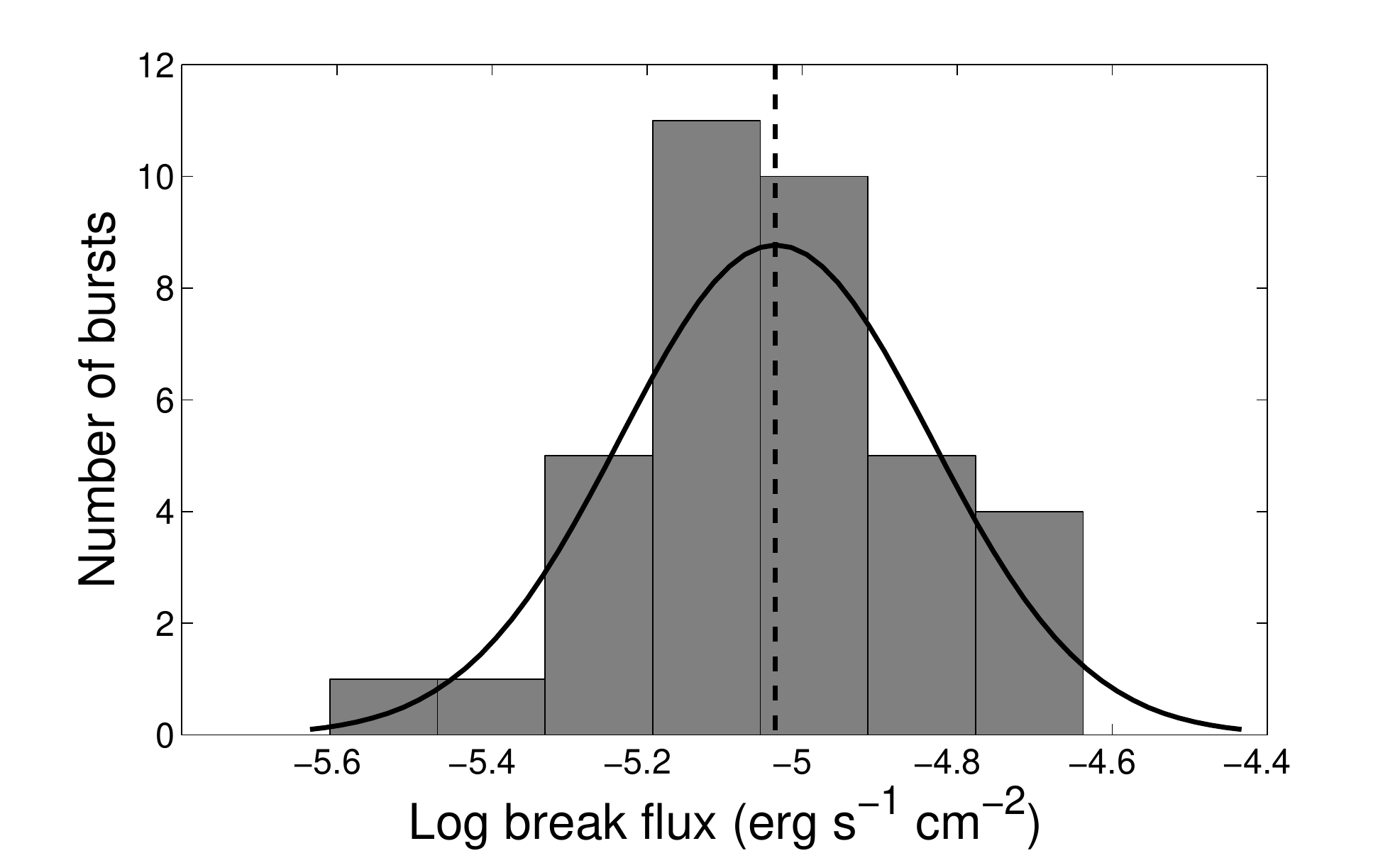}
\includegraphics[angle=0,width=0.48\textwidth]{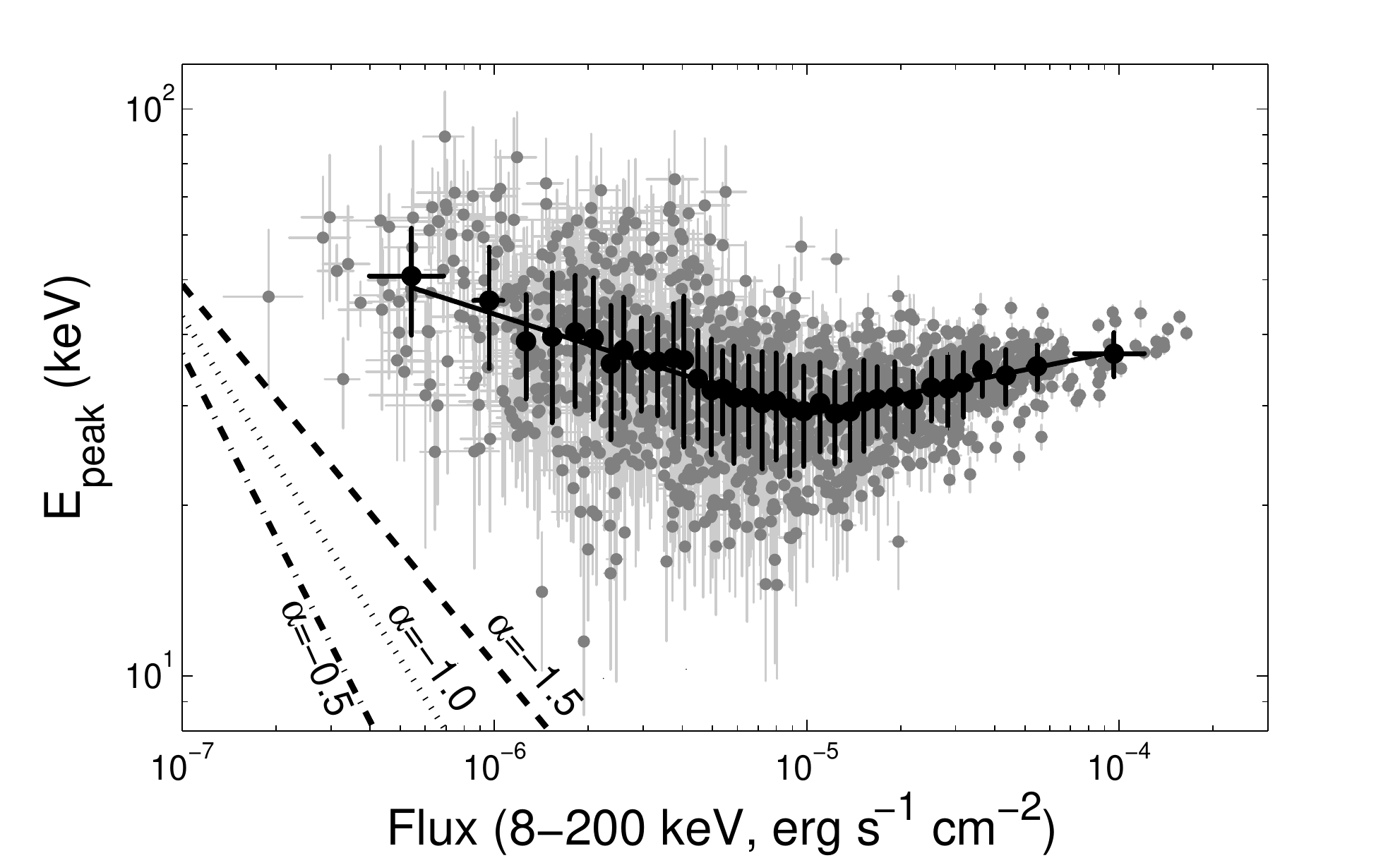}\\
\includegraphics[angle=0,width=0.48\textwidth]{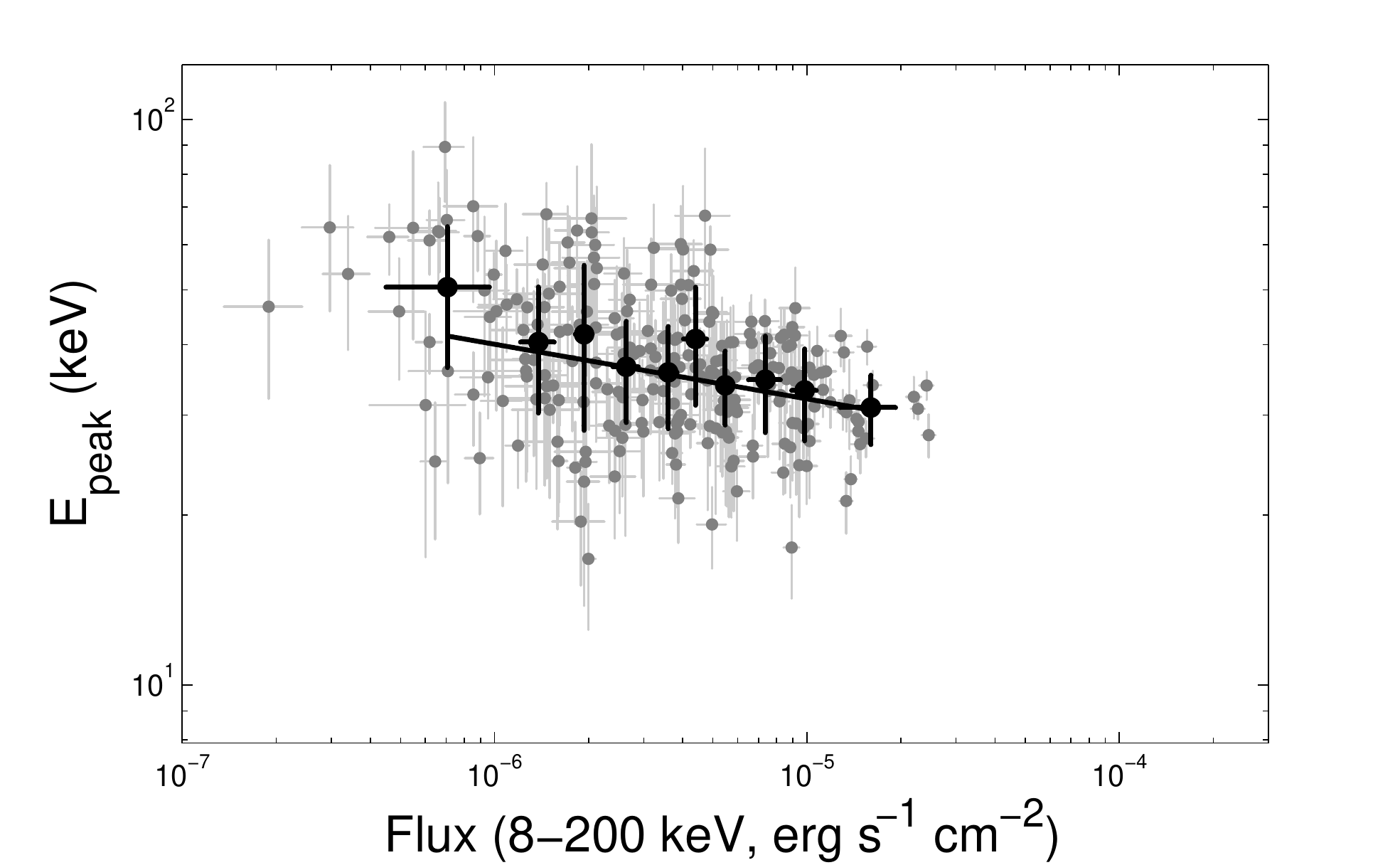}
\includegraphics[angle=0,width=0.48\textwidth]{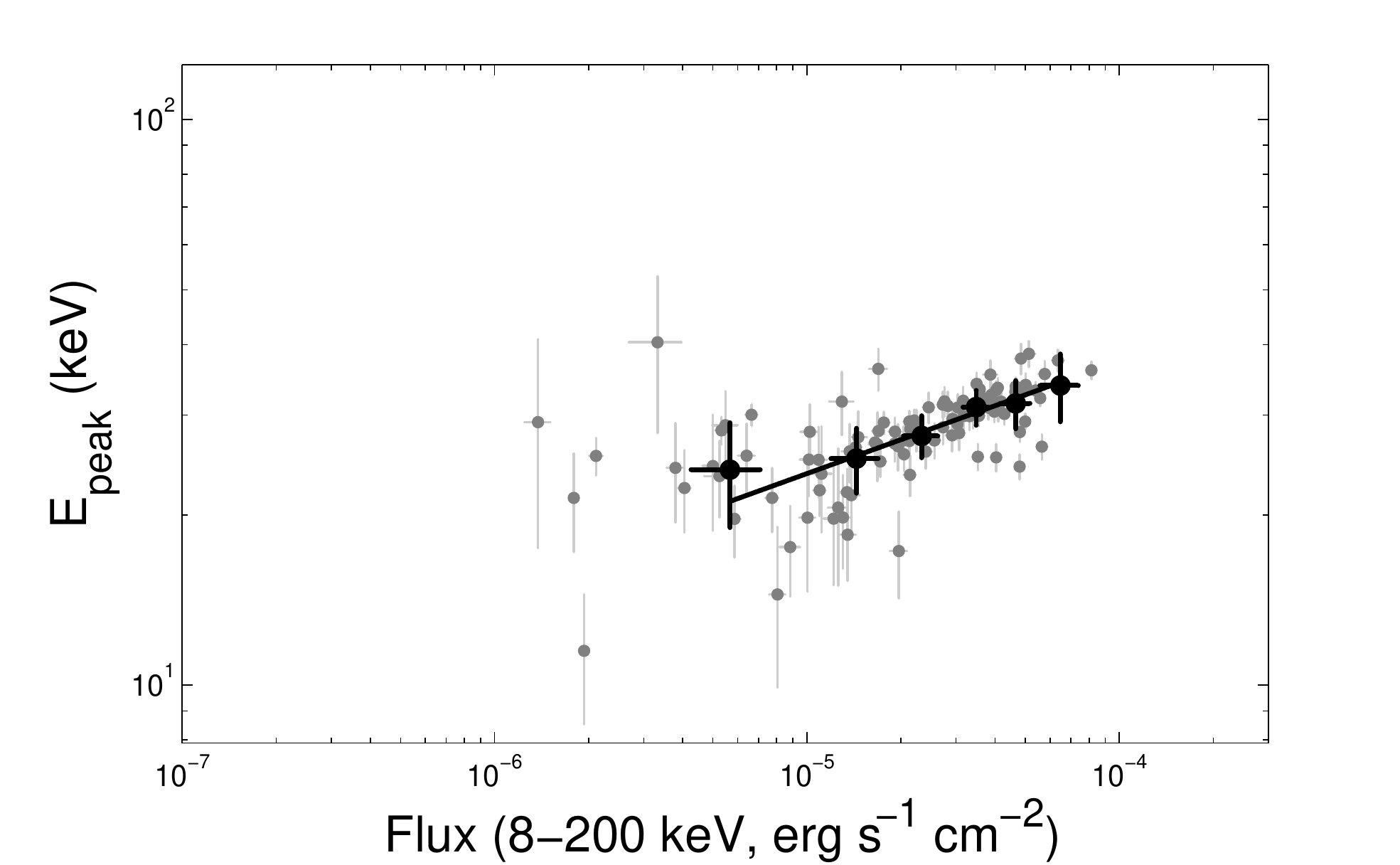}
\caption{{\sl Upper-left panel.} Distribution of the break flux values in the $E_{\rm
    peak}$ evolution as a function of flux for the 37 bursts in our
  sample showing a break in the correlation between these two
  parameters. The solid and dashed lines are the best fit Gaussian
  model to the distribution and its respective mean. {\sl Upper-right
    Panel.} \epeak\ evolution as a function of the 8-200 keV flux for
  all 49 bursts in our sample. Each dark-grey point is the result of a
  COMPT fit to a burst bin. The black dots
  represent the flux-binned data with 40 individual data points per
  bin. The black solid line is the best fit BPL model to the binned
  data. The dahsed, dotted, and dashed-dotted lines delimit the
  flux$-$\epeak\ parameter space, for index values of -1.5, -1.0, and 
  -0.5, respectively. {\sl Lower-left panel.} \epeak\ evolution as a
  function of the 8-200 keV flux for the 7 bursts with not enough
  data points above the flux break value. {\sl Lower-right panel.}
  \epeak\ evolution as a function of the 8-200 keV flux for the 5
  bursts with not enough data points below the flux break value. See
  text for more details.}
\label{breakVSflux}
\end{center}
\end{figure*}

\subsection{Comptonized model}
\label{compMod}

We started by fitting the COMPT model to all selected 4\,ms time
bins of each burst.  The fit parameters of some time bins within
the bursts were not well constrained due to a low number of counts.
In such cases we binned the data with a coarser time resolution to
accomplish a minimum constraint of 3$\sigma$ on $E_{\rm peak}$, the
peak energy of the COMPT spectrum.  As a result, 14 bursts have less
than 5 time bins for spectral fitting, and we removed these bursts
from our sample.  Our final sample for spectral analysis with the
COMPT model is, therefore, 49 bursts with a total number of 1393 time
bins,  for which we studied the correlations of the COMPT model
parameters ($E_{\rm peak}$ and power-law index) with the 8$-$200~keV
flux.

We first studied the spectral evolution with flux of each burst
separately, plotting $E_{\rm peak}$ as a function of flux.  To
identify the trend of the evolution, we fit this correlation with a PL
and a broken PL (BPL) model.  The choice for a BPL model is based on
the time-integrated spectroscopy results of \src\ (V12), and
time-resolved  spectroscopy of SGR~J0501+4516 \citep{lin11ApJ:0501}
which indicated a more complex relation between \epeak\ and flux than
a PL (e.g., see Figure~\ref{epkExa}).  An F-test showed an improvement
in the fit, with a $>$99.99\% non-chance occurrence, for 19 out of the
49 bursts. A more relaxed F-test criterion of $>$95\% results in 37
out of the 49 bursts being better represented by a BPL than a PL (see
e.g., Figure~\ref{epkExa}). Interestingly, the flux values at which
the break occurs for these 37 bursts seem to cluster around a common
value and does not depend on any of the burst parameters, e.g.,
$T_{\rm 90}$, average flux, etc. A Gaussian fit to the distribution of
these break flux-values results in $F\approx(1.0\pm0.2)\times10^{-5}$
erg~s$^{-1}$~cm$^{-2}$ (Figure~\ref{breakVSflux}, upper-left panel).

For the remaining 12 bursts, a single PL model was a better fit  to
the \epeak\ versus flux evolution.  However, examining these
bursts closely, we noticed that none of them had more than 5 data
points above (7 out of 12) or below (5 out of 12) the break flux level
of the majority of events.  Figure~\ref{breakVSflux} displays these
single PL trends of the 7 and 5 events in the lower-left and
lower-right panels, respectively.  It is unclear whether these events
would have followed the same break trend described above, had they had
more intense or faint data points, respectively. Finally, we studied
the correlation between \epeak\ and flux for all 49 bursts
simultaneously,  and we show this result as dark-grey points in the
upper-right panel of Figure~\ref{breakVSflux}. Here we binned the data
in flux requiring 40 individual data points in each bin, with the
last bin (at the highest fluxes) having a slightly lower number of
points. Each black dot in Figure~\ref{breakVSflux} represents the
weighted averages (in flux and \epeak) of these 40 individual
measurments, while the error bars correspond to their 1$\sigma$
deviations. Although not shown here, we also binned the data sorted by
\epeak, and tried different number of data points per bin (from 20 to
60), and we obtained consistent results.  A BPL fit to the binned data
is preferred at the $>99.99\%$ level over a single PL model. The BPL
fit has a low-flux PL index of $-0.18\pm0.02$ and a high-flux PL index
of $0.12\pm0.02$, with the break between the two PL regimes at a flux
of $(9.1\pm0.7)\times10^{-6}$ erg~s$^{-1}$~cm$^{-2}$ (8$-$200~keV) and
a break \epeak\ of $32\pm2$~keV, consistent with the 37-burst only
fit. Finally, our PL indices and the break flux and \epeak\ values are
similar to the values found in the time-integrated spectroscopy of
\src\ bursts, and also in the time-resolved spectroscopy of
SGR~J0501+4516 bursts \citep{lin11ApJ:0501}.  These results are
further discussed in Section~\ref{disc}.

\begin{figure}[th!]
\begin{center}
\includegraphics[angle=0,width=0.48\textwidth]{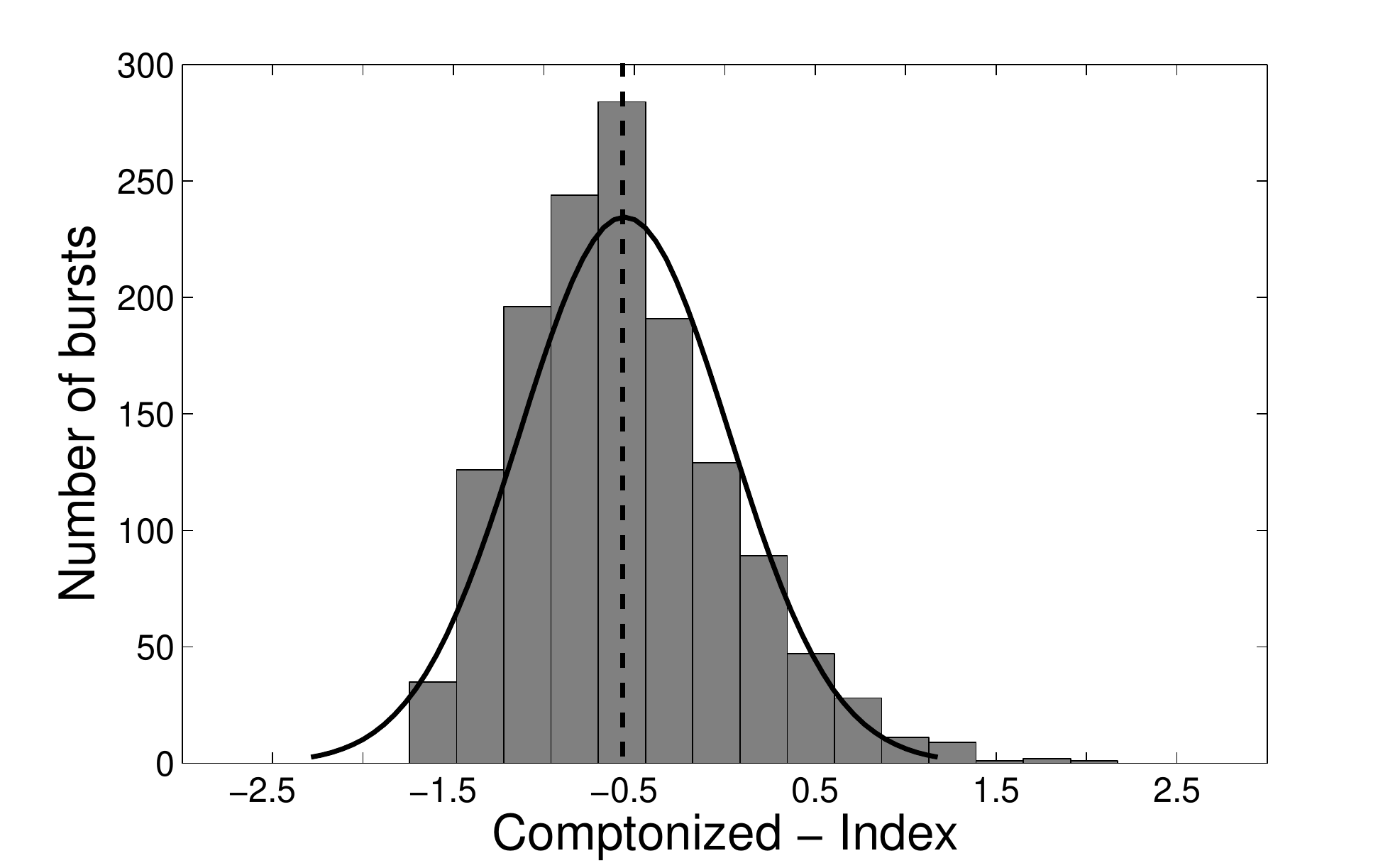}
\includegraphics[angle=0,width=0.48\textwidth]{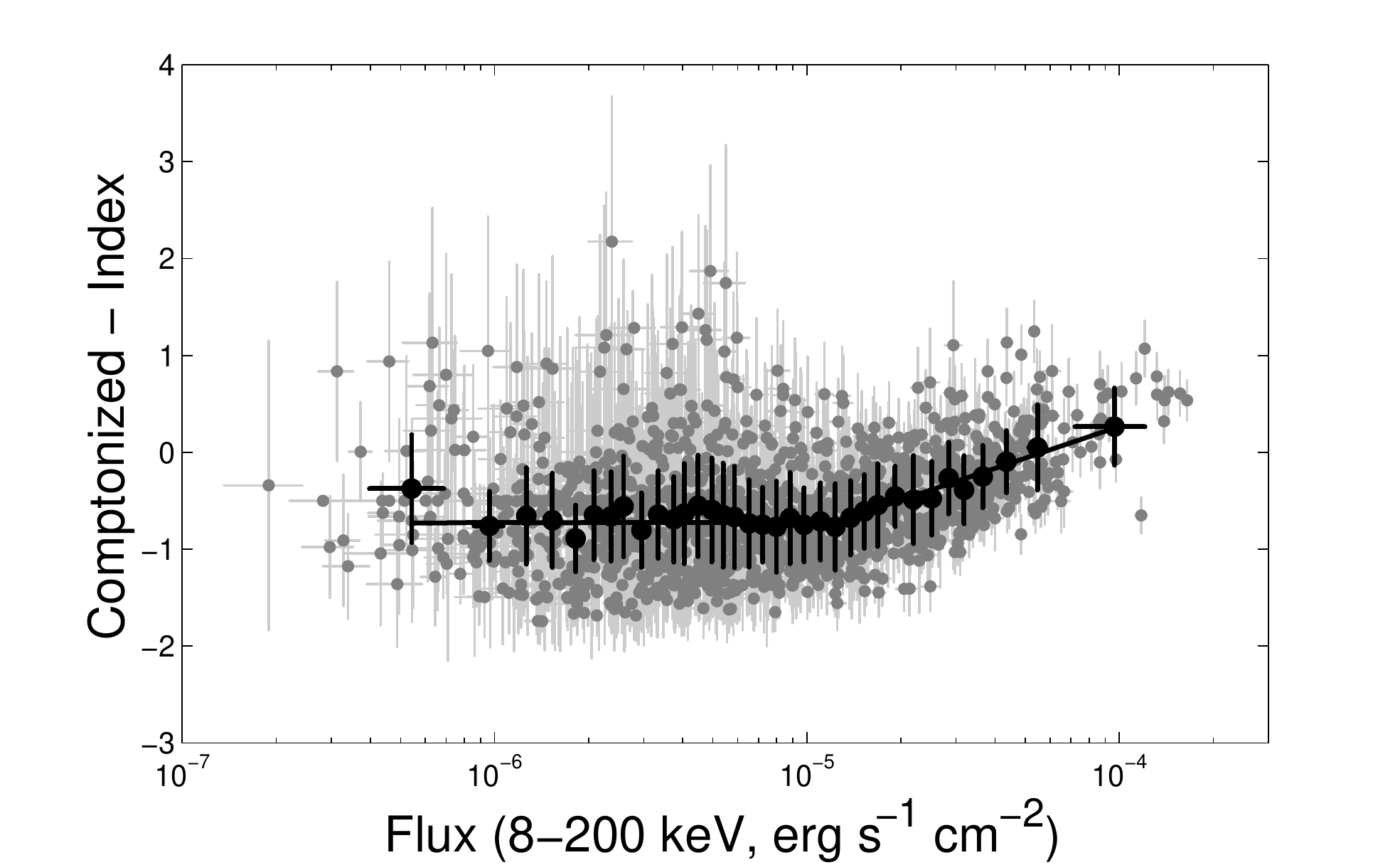}
\caption{{\sl Upper panel.} COMPT index distribution from our
  time-resolved analysis. The solid line is a Gaussian fit to the data
  and the dashed line represents the mean of the Gaussian
  $\approx-0.5$. {\sl Lower Panel.} COMPT index as a function of
  flux. Above the flux value of $10^{-5}$~erg~s$^{-1}$~cm$^{-2}$ the
  index positively correlates with flux, whereas below, it remains
  constant around $-0.8$.}
\label{indexVSflux}
\end{center}
\end{figure}

We find a more complicated correlation with flux than a single PL also
for the COMPT PL index. V12 have shown that the index of the COMPT
model in time-integrated spectroscopy is distributed narrowly around
$-1$.  Therefore, the OTTB model was an equally good fit to their
time-integrated spectra, since it is similar to a COMPT model with an
index of $-1$.  We show in the top panel of Figure~\ref{indexVSflux}
the distribution of the index for our time-resolved analysis.  The
distribution is broader than for the V12 time-integrated study and the
peak has now shifted to a higher value of $-0.55\pm0.58$. To
understand what prompted this shift, we plot in the bottom panel of
Figure~\ref{indexVSflux} the index as a function of flux.  Below the
flux value of $\sim10^{-5}$~erg~s$^{-1}$~cm$^{-2}$, identified as the
break point in the \epeak$-$flux correlation, the index appears
constant with flux, with a value around $\sim-0.8$. It is only above
the break where the indices become gradually steeper, following the
same trend observed between the \epeak\ and flux values. We also note
that, fitting some of the time-bins with the highest fluxes with the
OTTB model, results in a bad fit, as one would expect if the index
were deviating significantly from $-1$.

\subsection{Two blackbody (2BB) function}
\label{2BBmod}

We fit the initially selected 63 bursts (see Section~\ref{datSel})
with a 2BB model, using our 4~ms time-bins. We binned parts of the
light curves for significance, i.e., to different time
resolutions to accomplish a $3\sigma$ constraint on the low and high
BB temperatures. Since the 2BB model has more free parameters than the
COMPT one, coarser time bins were required.  Again, we removed bursts
from the initial sample with less than 5 time bins after re-binning,
resulting in 48 bursts and 994 individual time bins.

\begin{figure*}[th]
\begin{center}
\includegraphics[angle=0,width=0.48\textwidth]{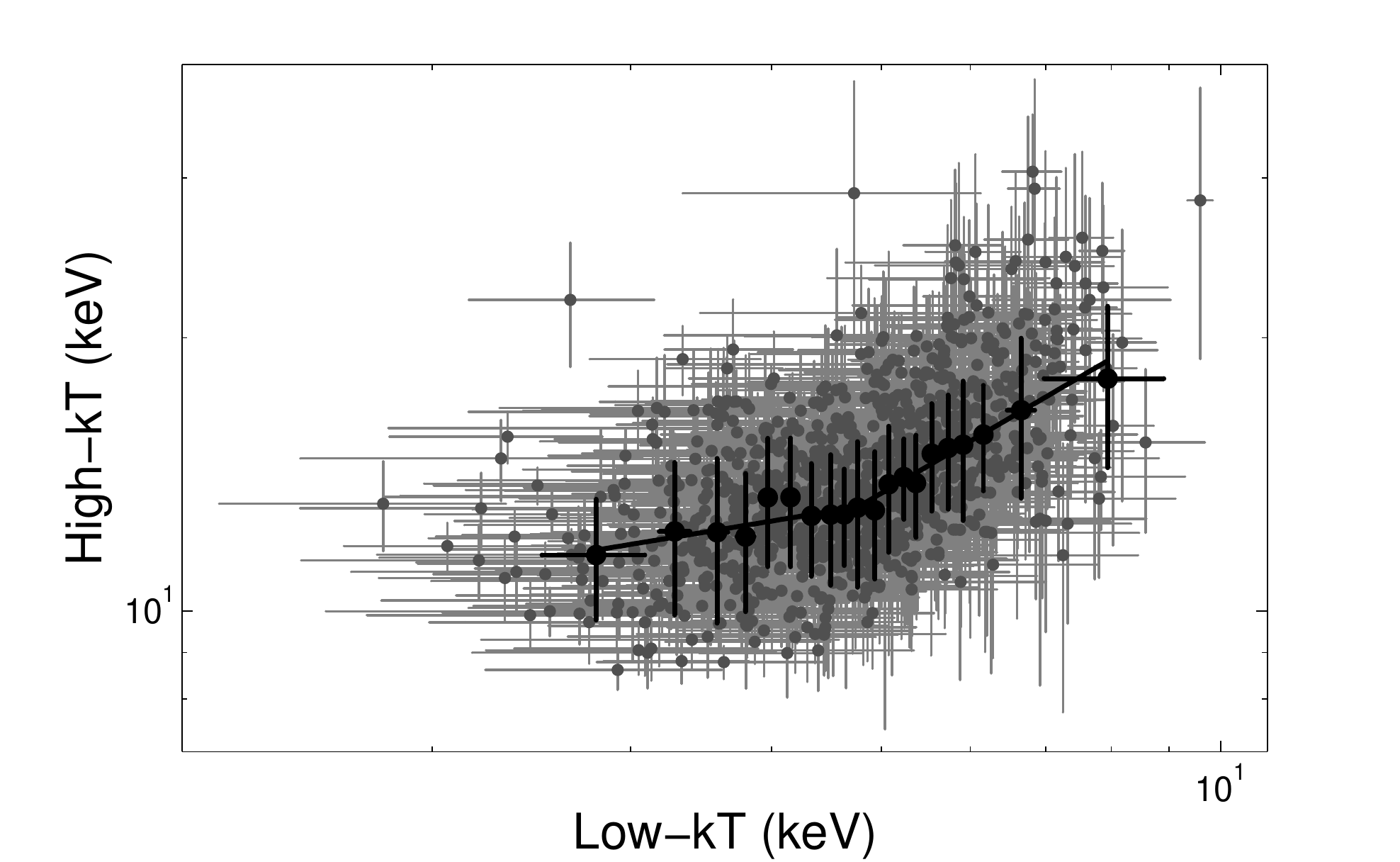}
\includegraphics[angle=0,width=0.48\textwidth]{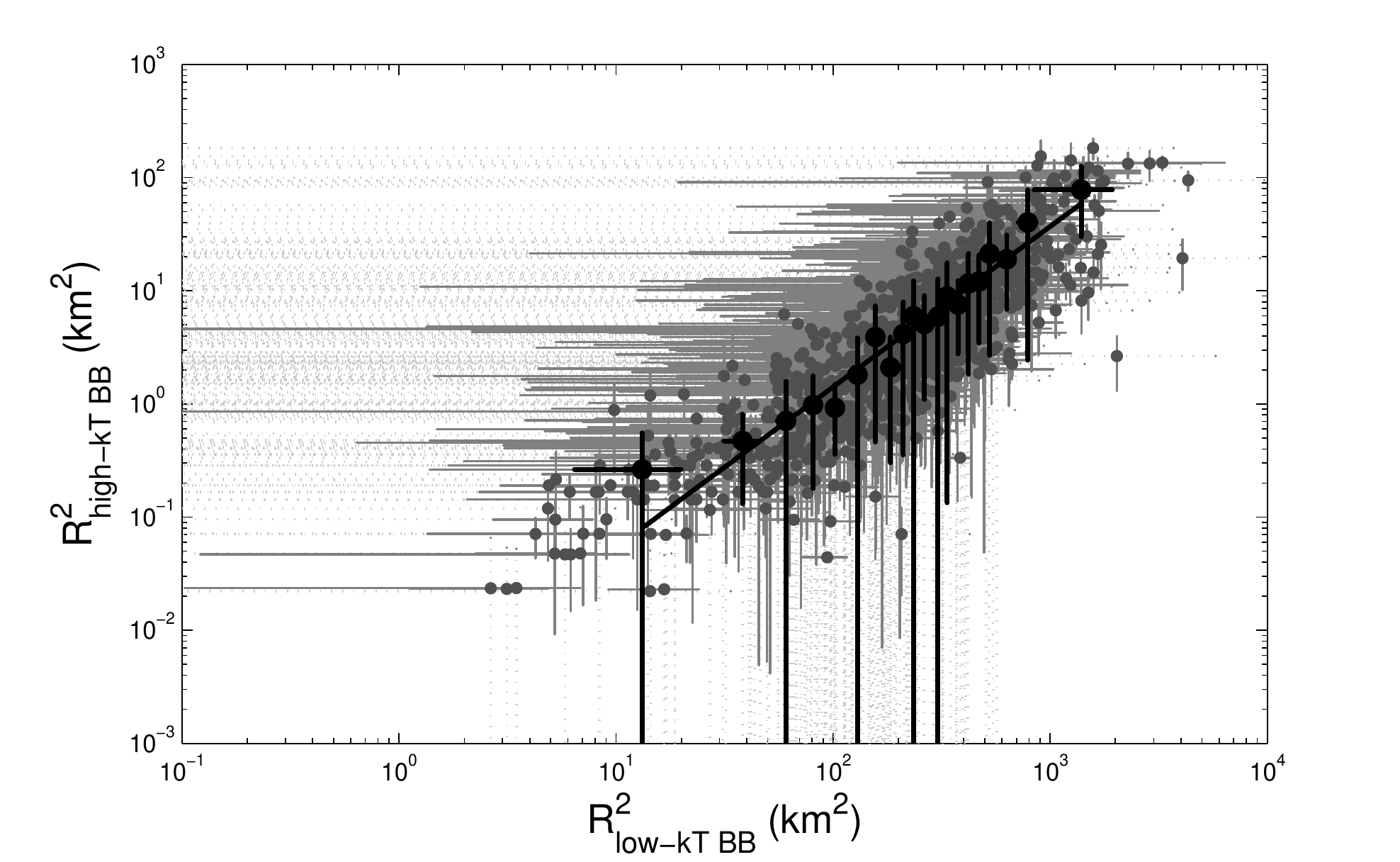}\\
\includegraphics[angle=0,width=0.48\textwidth]{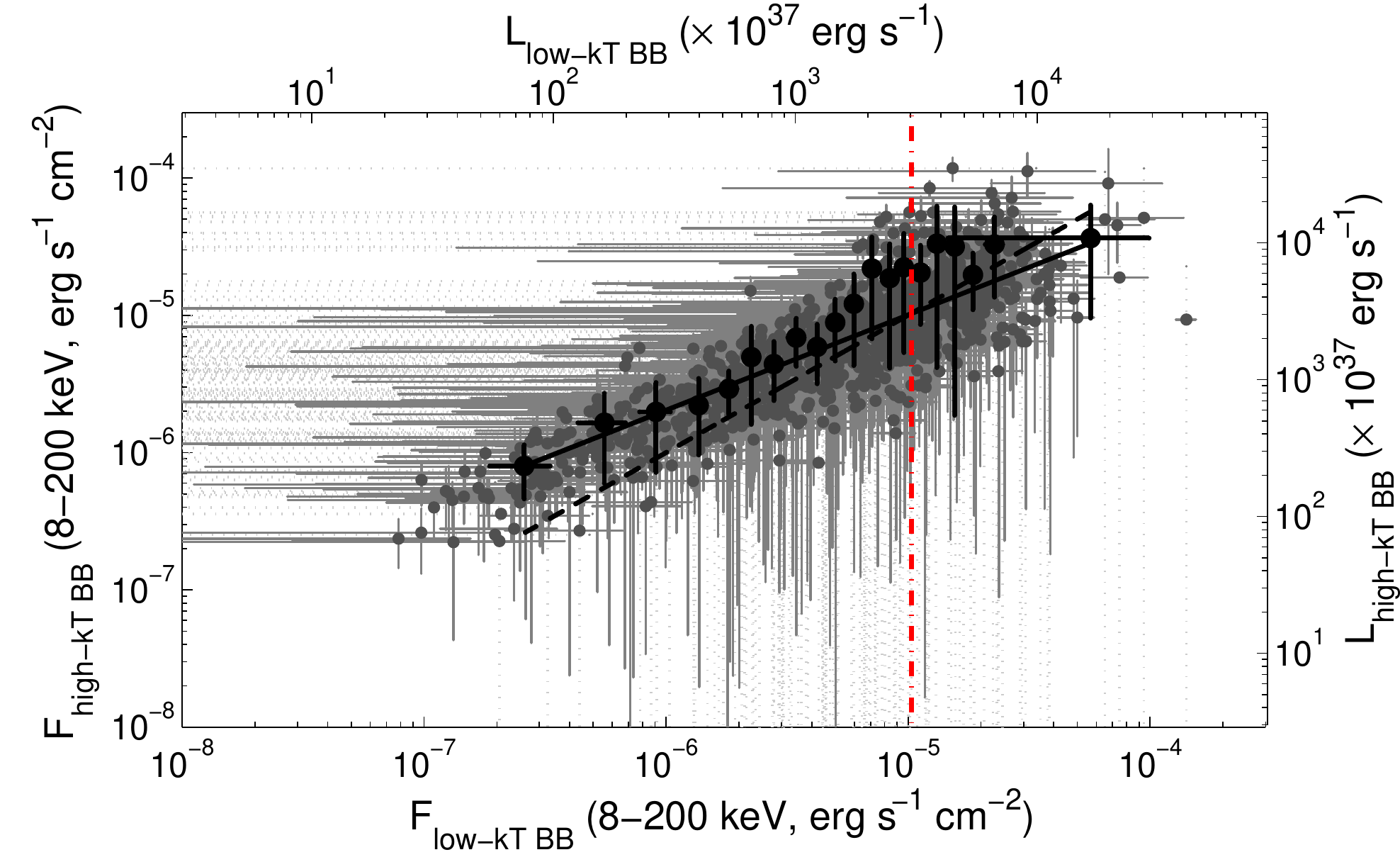}
\includegraphics[angle=0,width=0.48\textwidth]{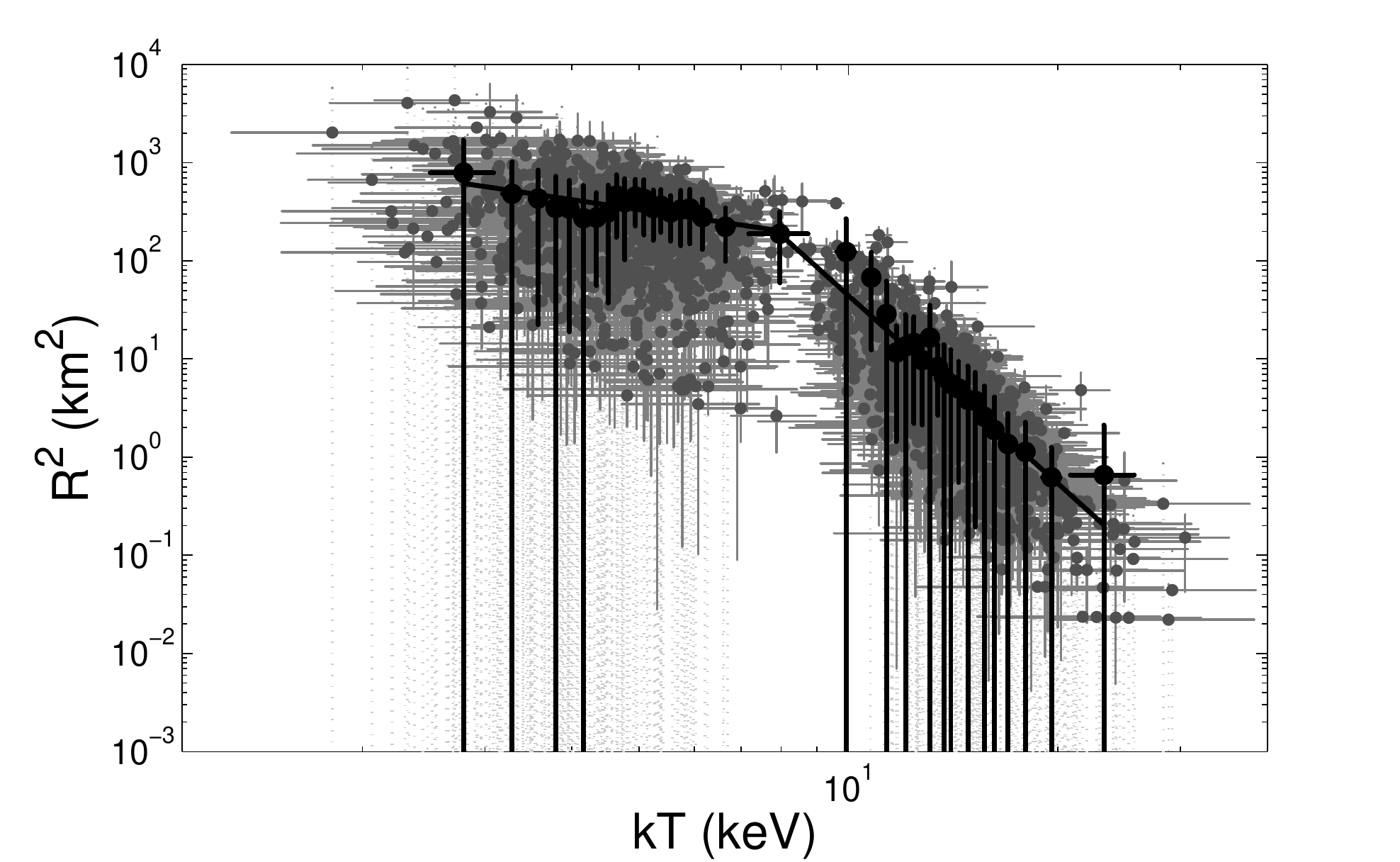}
\caption{{\sl Upper-left panel.} High$-kT$ versus low$-kT$ BB
  temperature. The solid line is a BPL fit to the data with a break at
  $\sim$4.8~keV. {\sl Upper-right panel.} Positive correlation between
  the radii of the emitting areas for the low- and high$-kT$ BB. {\sl
    Lower-left panel.} Positive correlation between the fluxes (and
  luminosities assuming a distance of 5~kpc) of the low- and high$-kT$
  BB. The dashed line represents a 1 to 1 relation. The dotted-dashed
  line is the $3\times10^{40}$~erg~s$^{-1}$ limit. {\sl Lower-right
    panel.} Radius of emitting area as a function of temperature for
  the low- and high$-kT$ BB simultaneously. The low- and high$-kT$ BB
  are clearly separated in the plot, with the area of the high$-kT$ BB
  decreasing at a faster pace compared to the area of the low$-kT$ BB,
  as a function of temperature. In all four panels, dark-grey dots
  represent individual measurements with grey crosses as their
  error bars; the light-grey dotted lines represent the errors of data
  points with uncertainty in the flux and/or area larger than its
  measurement value. Black dots are binned data with 50 points per
  bin. The black solid lines are the best fit PL or BPL model to the
  binned data.}
\label{ktvskt}
\end{center}
\end{figure*}

We examined several correlations between the fit parameters of the 2BB
model, i.e., (1) temperatures of the low temperature (low$-kT$)
BB versus high temperature (high$-kT$) BB; (2) the individual
fluxes of the 2 BBs; (3) the emitting areas corresponding to the
low$-kT$ and high$-kT$ BBs; (4) the radii of the emitting
areas versus temperature for both BBs simultaneously. The
effective radii of the emitting regions were calculated as in
\citet{lin11ApJ:0501}, $R^2=(FD^2/\sigma T^4)$~km$^2$, where $F$ is
the flux in erg~s$^{-1}$~cm$^{-2}$, $D$ is the distance to the source
in km (here assumed as 5~kpc, \citealt{gelfand07ApJ:1547,
  tiengo10ApJ:1547}), $\sigma$ is the Stefan-Boltzmann constant, and
$T$ is the temperature in Kelvin. We used the same approach as for the
COMPT model parameters to identify the trends of these different
correlations, fitting them for each burst separately to a PL and a BPL
model, and performing F-tests to evaluate the improvement in the
fits.

For 26 of the 48 bursts, we find that the temperatures of the
low$-kT$ BB versus high$-kT$ BB are best fit with a BPL ($>$95\%
probability for an improvement from a PL fit to occur by chance) with
indices and breaks consistent with each other. The rest of the bursts
follow similar but not statistically significant trends, due to low
number of data points. Figure~\ref{ktvskt}, upper-left panel, shows
the relation between low$-kT$ and high$-kT$ values for all 48 bursts
simultaneously. Here (and in all panels of Figure~\ref{ktvskt}), we
binned the data at 50 points per bin (black dots). We find that a BPL
is a better fit to the data (black-solid line) compared to a single
PL, at the $>99.99\%$ confidence level. Above the break, the relation
is much steeper than below the break with slopes of $0.7\pm0.2$ and
$0.2\pm0.1$, respectively. The corresponding break is at ($4.8\pm0.3,
13.1\pm0.4$)~keV. These results are different than the time-integrated
spectroscopy of these bursts, which showed only a positive correlation
between the low- and high$-kT$s, with a very weak hint for a possible
break at low temperatures (V12); in that work, the slope of the single
PL fit was much steeper, i.e., $1.86\pm0.09$.

To rule out any systematic effect as the origin of the break, and in
particular the fact that the temperature of the low$-kT$ BB falls
below the \fermi/GBM energy coverage, we looked at
\citet{lin12ApJ:1550}, who studied the combined spectra of the \src\
bursts simultaneously seen with \swift/XRT and GBM, thus extending the
energy coverage down to 1~keV (see also Section~\ref{simu}). The
authors found that the low- and high$-kT$ temperatures derived solely
with GBM are equal to the ones derived from the joint XRT/GBM
fits. The only deviations ($<1\sigma$) between the two results are
noticeable at very low temperatures (below 4.8 keV for the low$-kT$),
where the GBM-only fits gave slightly (but consistently) higher
low$-kT$ temperatures than the joint ones. Therefore, if the break we
see in our relation is due to systematic effects, one would also
expect another break below a certain low$-kT$, where the relation
becomes steeper.

\begin{figure}[th!]
\begin{center}
\includegraphics[angle=0,width=0.48\textwidth]{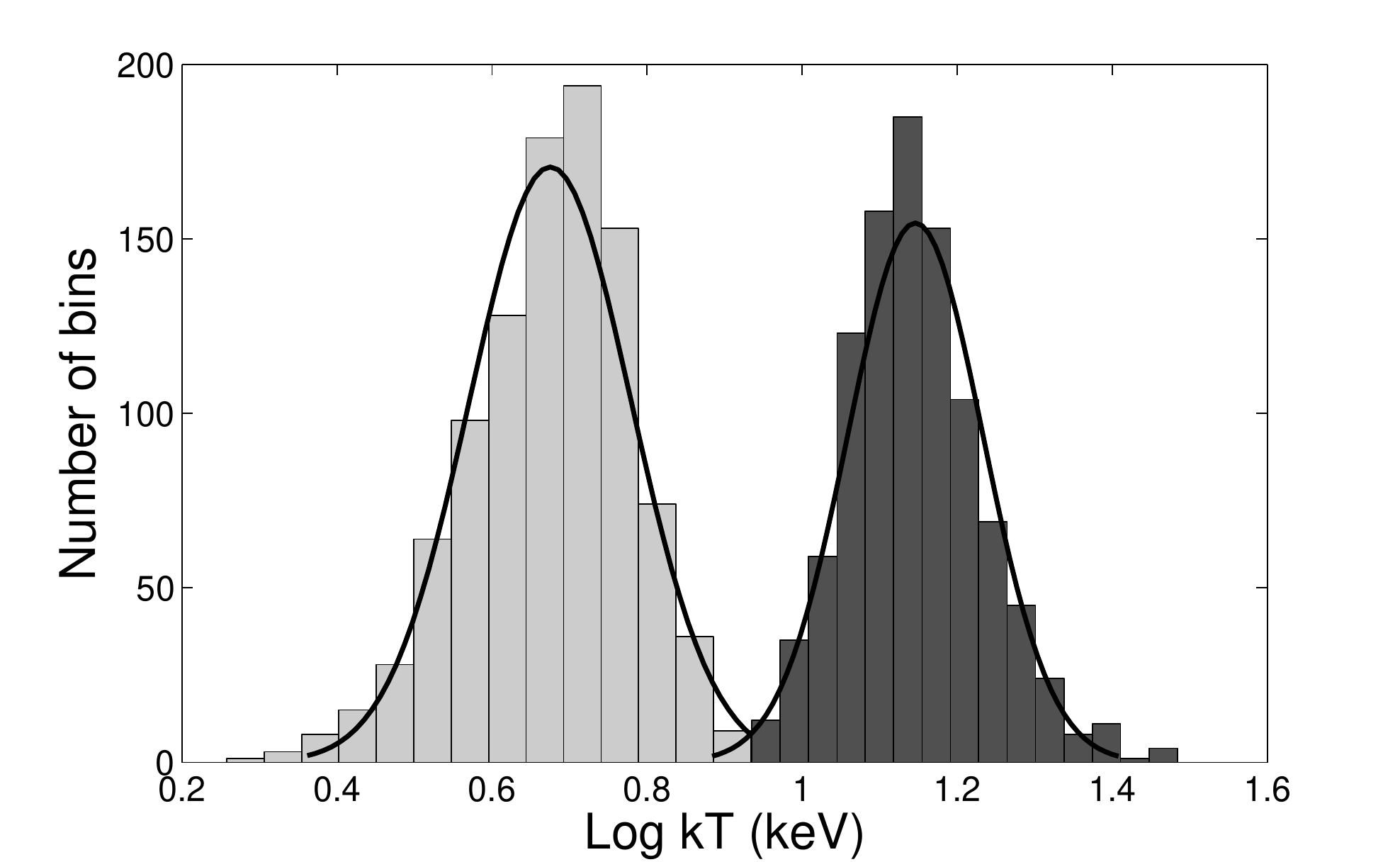}
\caption{Distributions of the low- and high$-kT$ BB temperatures with their respective
  best fit Gaussian models (solid lines).}
\label{ktdist}
\end{center}
\end{figure}

The areas of the emitting regions as well as the fluxes of the low-
and high$-kT$ BBs are positively correlated, with consistent single PL
slopes for all 48 bursts in our sample. These correlations are shown
at the upper-right and lower-left panels of Figure~\ref{ktvskt} (the
areas are represented as $R^2$, with $R$ the radius).  The PL slope of
the correlation between the areas is $1.2\pm0.3$, while for the fluxes
it is $0.7\pm0.1$.  These slopes are consistent with the slopes for
the time-integrated analysis of \src\ GBM bursts, which were
$1.34\pm0.04$ and $0.83\pm0.02$, respectively. The slope of the
correlation between the low- and high$-kT$ BB fluxes is close to
the 1 to 1 relation (dashed line), indicating that the BBs contribute
almost equally to the total burst flux. We note that in the case of
time-resolved spectral analysis of SGR~J0501+4516 bursts this slope is
$1.1\pm0.1$ \citep{lin11ApJ:0501}, while for SGR~1900+14 it is
$0.70\pm0.03$ \citep{israel08ApJ:1900}.

Besides comparisons between the two BB components, we investigated the
relation between the emission area and temperature, for both the low-
and high$-kT$ BB simultaneously. Most of the bursts showed a similar
trend between these two parameters, namely that the area of the
low$-kT$ BB decreases with temperature at a different pace than the
area of the high$-kT$ BB. This same trend is seen when combining all
the individual time bins, as shown in the lower-right panel of
Figure~\ref{ktvskt}.  A BPL fit to the binned data has slopes of
$-1.0\pm0.3$ and $-6.4\pm0.5$ for the low$-kT$ and high$-kT$
correlations, respectively and a break at a temperature of $\sim7.9$~keV.

To determine the significance of this last relation, we calculated the
Spearman rank correlation coefficients of the low- and high$-kT$ BB
areas versus their corresponding temperatures. We found a correlation
coefficient $r\approx-0.7$ and a probability of $\sim99.9\%$ that the
low$-kT$ area is correlated to the temperature, while for the
high$-kT$ BB we found a stronger correlation $r\approx-1.0$ and a
probability $>99.99\%$. The time-integrated spectroscopy of the \src\
bursts (V12) hint at a similar relation for the high$-kT$ BB, however,
the low$-kT$ BB did not show any correlation between area and
temperature.

\begin{figure*}[t]
\begin{center}
\includegraphics[angle=0,width=0.48\textwidth]{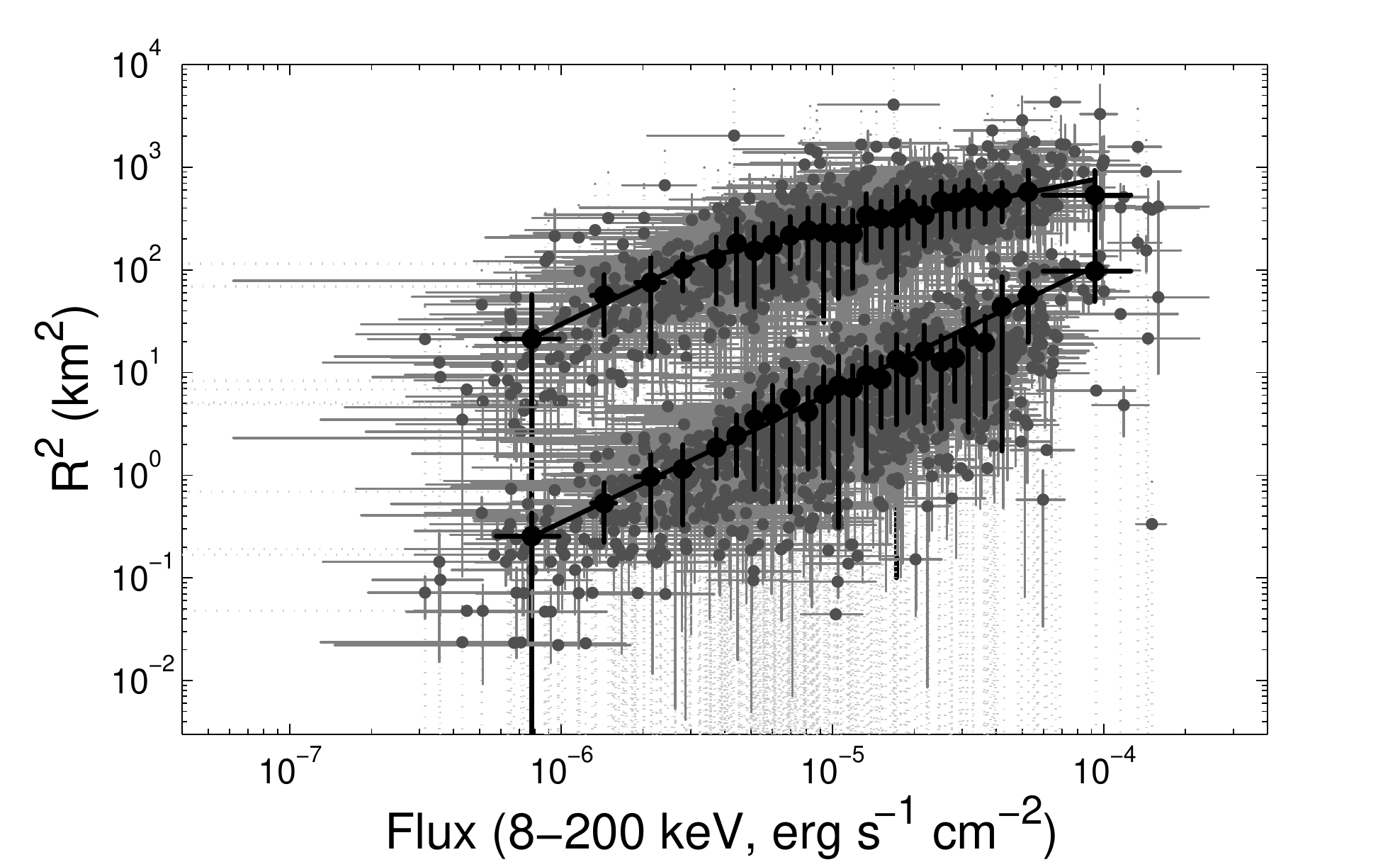}
\includegraphics[angle=0,width=0.48\textwidth]{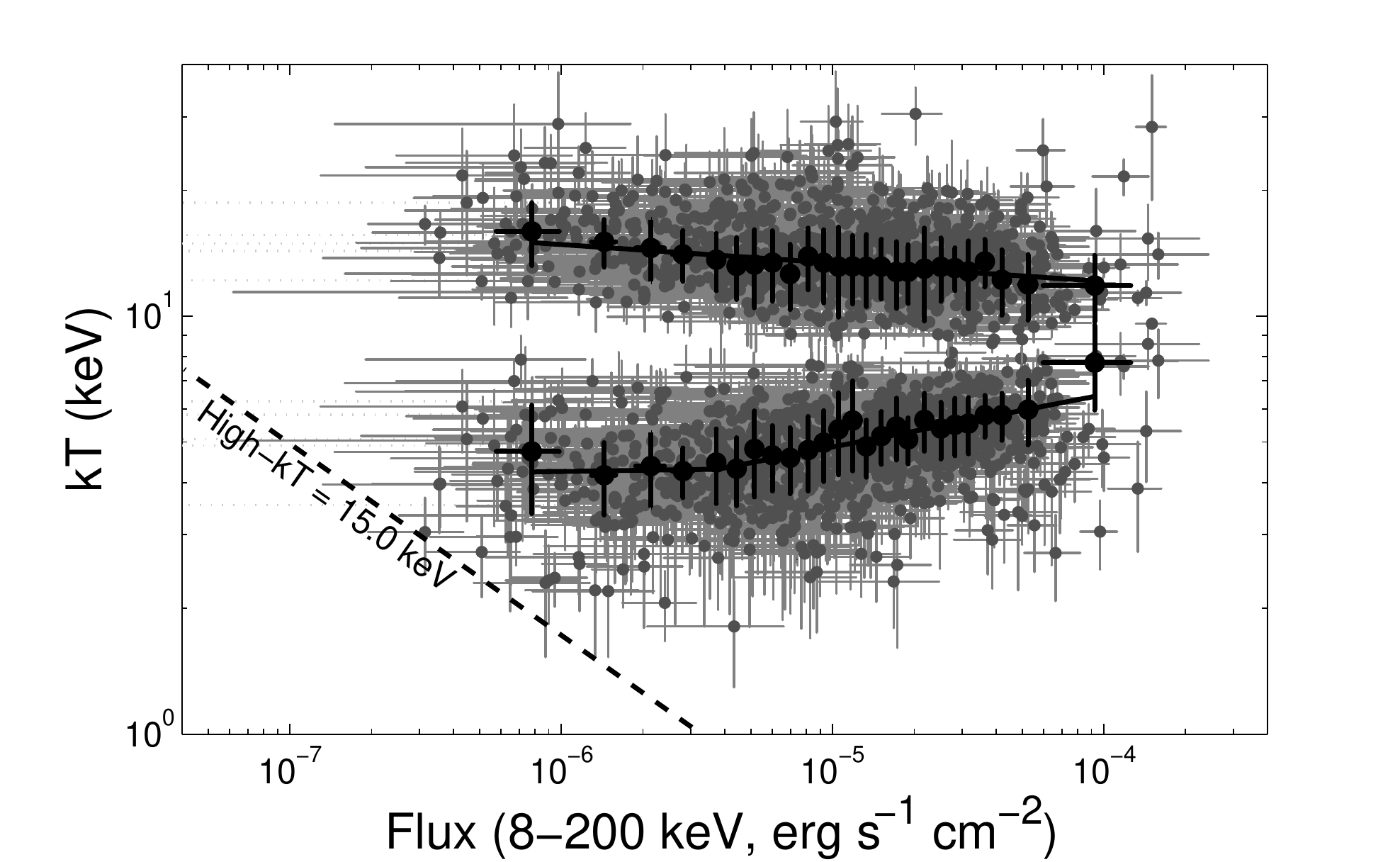}\\
\includegraphics[angle=0,width=0.48\textwidth]{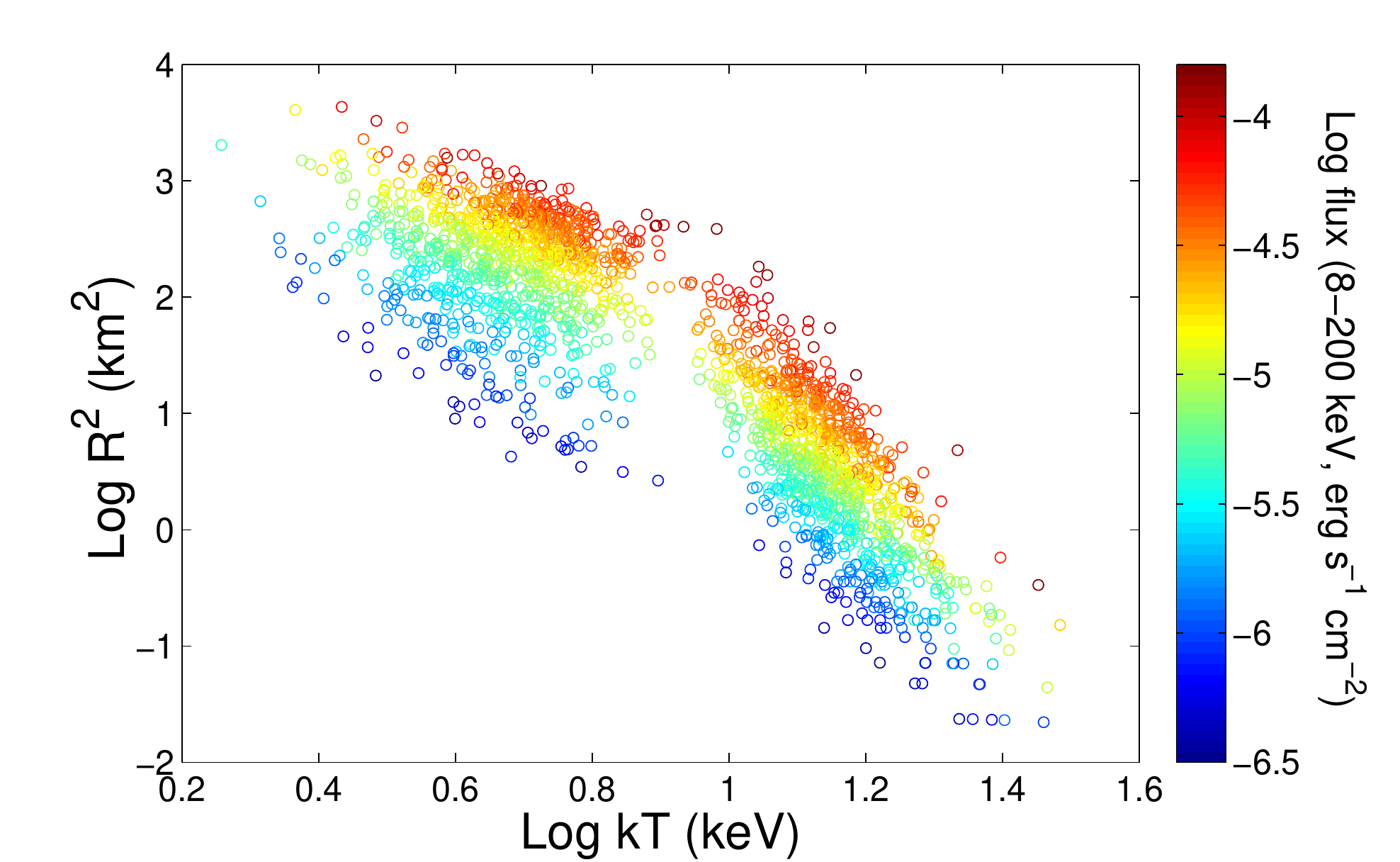}
\includegraphics[angle=0,width=0.48\textwidth]{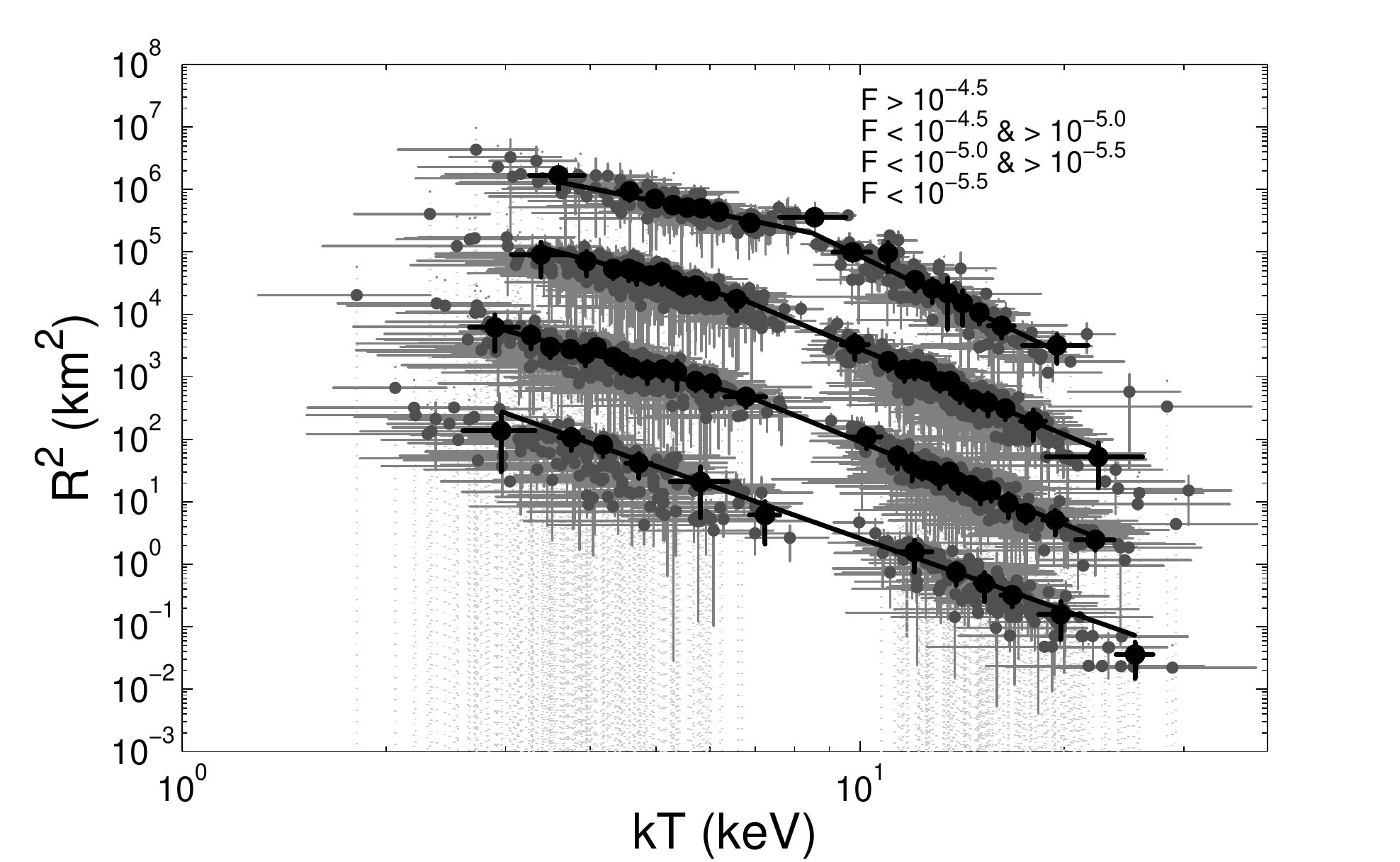}
\caption{{\sl Upper-left panel.} Low- and high$-kT$ BB areas as a function
  of flux. {\sl Upper-right panel.} Low- and high$-kT$ BB temperatures as a
  function of flux. The dashed line delimit the flux$-$low-kT parameter
  space for a high-kT value of 15~keV. {\sl Lower-left panel.} Flux
  color-coded plot of the area as a function of temperature. {\sl
    Lower-right panel.} Area as a function of temperature grouped in
  flux intervals from top to bottom as indicated by the legend. Flux
  groups are shifted arbitrarily for clarity. See text for details.}
\label{2BBvsFlux}
\end{center}
\end{figure*}

Finally, we investigated the evolution of temperature and area with
flux, $F$. The upper-left panel of Figure~\ref{2BBvsFlux} (here we
binned the data at 40 points per bin, i.e., black dots) shows the
relation between the low- and high$-kT$ areas (corresponding to the
high- and low$-kT$ BB temperatures) and fluxes.  We again fit these
correlations with a PL and a BPL to look for any possible breaks. It
is clear that the area of both the low and high temperature BB
increases with increasing flux.  There is, however, a tendency for the
low$-kT$ area to flatten at the highest fluxes, while the high$-kT$
area increases steadily. Indeed, the low$-kT$ area as a function of
flux is best fit with a BPL with slopes of $1.3\pm0.2$ and
$0.5\pm0.1$, respectively, and a break at
$F=(3\pm1)\times10^{-6}$~erg~s$^{-1}$~cm$^{-2}$, whereas the high$-kT$
area is best fit with a single PL with a slope of $1.3\pm0.1$. A
Spearman rank test shows  that the correlation coefficients and
probabilities for the low- and high$-kT$ areas as a function of flux
are $r=0.8$, $p>99.99\%$ and $r=0.7$, $p>99.99\%$, respectively. We
estimate that about $92\%$ of the low$-kT$-area data points with
$F>10^{-4.5}$~erg~s$^{-1}$~cm$^{-2}$ land at $\lesssim10^3$~km$^2$,
which corresponds to the 1$\sigma$ limit of the highest-flux binned
data point.

The temperatures of the 2 BBs follow opposite trends with respect to
flux, as can be seen in the upper-right panel of
Figure~\ref{2BBvsFlux}.  The low$-kT$ appears to positively correlate
with flux, only above a break limit. This relation is best fit with a
BPL, with slopes of $0.0\pm0.1$ and $0.13\pm0.02$, below and above
$F=4\pm2)\times10^{-6}$~erg~s$^{-1}$~cm$^{-2}$. The high$-kT$, on the
other hand, shows a mild negative correlation with flux, and it is
best fit with a single PL with a slope of $-0.04\pm0.02$. The
Spearman-rank correlation coefficients and probabilities are  $r=0.5$,
$p>99.99\%$, and $r=-0.3$, $p>99.99\%$, for the low- and high$-kT$ as
a function of flux, respectively.

Given the correlations between the area and temperature with flux,
next we examined the effect of the flux on the relation between area
and $kT$. We created a color-coded plot by flux of these two
parameters, which is shown in the lower-left panel of
Figure~\ref{2BBvsFlux}.  We note that the relation between area and
$kT$ remains the same across all flux values, decreasing more steeply
for the high temperatures compared to low temperatures.  We show that
also by separating our data into four flux ranges ($F <10^{-5.5}$, 
$10^{-5.5}<F <10^{-5.0}$, $10^{-5.0}<F <10^{-4.5}$ and $F>10^{-4.5}$),
and fitting a PL and a BPL to the data in each flux range separately
(Figure~\ref{2BBvsFlux}, lower-right panel; flux groups have been
shifted for clarity).  For the three highest flux ranges we find that
a BPL is preferred over a PL fit at the $>99.99\%$ level, while for $F
<10^{-5.5}$ a BPL is only required by the data at the $90\%$
level. Table~\ref{areakTfluxes} shows all fit results.

\begin{table}[!t]
\caption{Area ($R^2$) versus $kT$ fit parameters for the different
  flux groups in the bottom-right panel of Figure~\ref{2BBvsFlux}.}
\label{areakTfluxes}
\newcommand\T{\rule{0pt}{2.6ex}}
\newcommand\B{\rule[-1.2ex]{0pt}{0pt}}
\begin{center}{
\resizebox{0.5\textwidth}{!}{
\begin{tabular}{l c c c}
\hline
\hline
Flux range \T \B & Slope below $kT_{\rm break}$ & Slope above $kT_{\rm
  break}$ & $kT_{\rm break}$ \\
erg s$^{-1}$ cm$^{-2}$ \T \B &  &  & keV \\
\hline
$F>10^{-4.5}$ \T               & $-2.2\pm0.3$ & $-5.3\pm0.4$ & $9\pm1$ \\
$10^{-5.0}<F<10^{-4.5}$ & $-2.8\pm0.3$ & $-4.6\pm0.4$ & $7\pm1$   \\
$10^{-5.5}<F<10^{-5.0}$ & $-3.0\pm0.3$ & $-4.4\pm0.5$ &  $7\pm1$  \\
$F<10^{-5.5}$                &  \multicolumn{3}{c}{$-3.8\pm0.4$$^*$}    \\
\hline
\end{tabular}}}
\end{center}
\begin{list}{}{}
\item[{\bf Notes.}]$^{(*)}$ A single PL fit to the data.\\
\end{list}

\end{table}

\subsection{Simulations}
\label{simu}

We performed simulations with RMFIT (rmfit4.3) to determine the
parameter space for the COMPT and 2BB models, for a typical magnetar
burst spectrum detected with GBM.

We first checked whether we can recover the different spectral
parameters we get from the real data through simulations of a large
set of synthetic spectra. We chose three bursts with at least 10
bins at low fluxes ($<5\times10^{-6}$~erg~s$^{-1}$~cm$^{-2}$). We then
generated 1,000 synthetic spectra for each bin and each relevant
detector. Each synthetic spectrum was based on the sum of the
predicted source counts and the measured background counts in each
energy channel. The former counts were computed from the analytical
function used to fit the real data (COMPT or 2BB) folded with the
instrumental response function of the relevant detector. The
background counts were estimated for each detector from the real
data. Next, we applied Poisson fluctuations to the summed counts to
obtain the final synthetic spectrum (see also, V12 and \citealt{
  guiriec11ApJ:grb724}). The resulting synthetic spectra were fitted
with the model used to generate them. All input spectral parameters of
both models were well recovered in these fits by comparing them and
their statistical errors to the simulated parameter distributions and
their uncertainties.

Next, we generated 1000 synthetic spectra using random input spectral
parameters for the COMPT and 2BB models, decreasing the 8-200~keV
flux from $4\times10^{-6}$~erg~s$^{-1}$~cm$^{-2}$ down to
$10^{-7}$~erg~s$^{-1}$~cm$^{-2}$, in steps of 0.2 dex. We then checked
if the input spectral parameters were recovered within 1$\sigma$ , at
the $90\%$ confidence level.

For the COMPT model, we varied the index between three different
values ($\alpha=-0.5, -1.0, -1.5$), in order to have a better handle
on $E_{\rm peak}$. The COMPT parameter space is shown in the
upper-right panel of Figure~\ref{breakVSflux}, for all three values of
the index; for each value, the area below the line cannot be retrieved
with the instrumental sensitivity of GBM. It is clear from the
simulations that even with a soft spectrum ($\alpha=-1.5$), we should
have been able to detect bursts with $E_{\rm peak}$ as low as 10 keV
for a 8-200~keV flux of $\sim10^{-6}$~erg~s$^{-1}$~cm$^{-2}$. The flux
level goes down to $3\times\sim10^{-7}$~erg~s$^{-1}$~cm$^{-2}$ for a
burst with a hard spectrum, $\alpha=-0.5$. Hence, we can conclude that
the break we see in the $E_{\rm peak}-$flux (and $\alpha-$flux)
relation at $\sim10^{-5}$~erg~s$^{-1}$~cm$^{-2}$ is not due to any
systematic effects. We note that a very soft burst with $\alpha=-2.0$,
would have been detected down to a $E_{peak}\approx16$~keV, had it had
a flux of $\sim10^{-6}$~erg~s$^{-1}$~cm$^{-2}$.

For the 2BB model, and similar to COMPT, we varied high-kT in a
pre-defined set of values of 10, 15, and 20 keV, in order to have a good
handle over the low-kT spectral parameters, which are harder to
determine with GBM. We made sure to vary the normalizations of the two
components in a way that both BBs contribute equally to the total
flux, according to the well established relation between the low- and
high-kT fluxes (Figure~\ref{ktvskt}, \citealt{israel08ApJ:1900}). We
first found that the temperature of the high-kT component has little
effect on the constraint of the low-kT spectral parameters. In the
upper-right panel of Figure~\ref{2BBvsFlux}, we show the parameter
space of the low-kT temperature with total flux. The parameter space
in case of the 2BB model is smaller than in the case of COMPT, which
is expected considering the extra free parameter. Nonetheless,
according to our simulations, we should have been able to detect
bursts with temperatures falling in the area of the extrapolation of
the low-kT relation with flux seen above the break, hence concluding
that the break we see in the $kT-$flux relation is real, and not due
to systematic effects.

\section{Discussion}
\label{disc}

\subsection{Comparison to previous results}

A large number of bursts were emitted by \src\ during the 2008-2009
active episode, especially on 2009 January 22 when the source emitted 
hundreds of bursts in nearly 24 hours. These bursts were detected by a
multitude of high-energy space instruments. \citet[][see
also \citealt{   mereghetti09ApJ:1547}]{savchenko10AA:1547} analyzed
about 100 bursts detected on 2009 January 22  with the {\sl INTEGRAL}
SPI$-$ACS and ISGRI detectors. \citet{scholz11ApJ:1547} presented a
detailed time-averaged spectral study of hundreds of bursts from \src\
detected with \swift/XRT. Finally, \citet{enoto12MNRAS:1550} studied 13
very weak bursts detected with {\sl Suzaku} from \src\
($F\lesssim10^{-8}$~erg~cm$^{-2}$). The better energy coverage and
sensitivity of \fermi/GBM compared to these instruments allowed for a
more detailed time-averaged spectroscopy of hundreds of bursts from
\src\ (V12, see also \citealt{vonkienlin12ApJ:1550}). A COMPT model
and a 2BB model were used to fit the time-integrated spectra, and V12
looked for correlations between the different fit parameters. For the
COMPT model, V12 found a hint for a weak double correlation between
\epeak\ and flux as opposed to the strong one we find here
(Figure~\ref{breakVSflux}). This could be attributed to the fact that
only very few bursts showed an average flux above the flux break-limit
that we establish here (i.e.,
$F\sim10^{-5}$~erg~s$^{-1}$~cm$^{-2}$). On the other hand, for the 2BB
model, V12 found a strong positive correlation between the low$-kT$
fluence and high$-kT$ fluence with a slope of 0.83, and between the
low$-kT$ area and the high$-kT$ area with a slope of 1.34. These
results are in perfect agreement with the results we derive for our
time-resolved spectroscopy (Figure~\ref{ktvskt}, but with fluxes
instead of fluences). However,  the area$-kT$ relation is clearer in
our analysis than in V12, where we show that the low$-kT$ area
decreases with temperature, only at a slower pace than the high$-kT$
area. This could be due to a combination of better statistics (more
data) when performing time-resolved spectroscopy and/or the fact that
the relation might be smeared out when performing time-integrated
spectroscopy. \citet{lin12ApJ:1550} studied the bursts of \src\ seen
simultaneously with \swift/XRT and GBM, hence, broadening the energy
coverage down to $\sim1$~keV. Interestingly, fitting the time-integrated
spectra of these bursts with the 2BB model, \citet{lin12ApJ:1550}
recovered, although at a much lower significance (most likely due to
much lower statistics), the trend we see here between the areas and
temperatures of the 2 BBs (Figure~\ref{ktvskt}, lower-right panel). We
also note that the values of our fit parameters are comparable to
results derived from time-integrated spectroscopy of the bursts of
other sources, {\it e.g.}, SGR $1900+14$, SGR $1806-20$, and SGR J$0501+4516$
\citep{feroci04ApJ:1900,nakagawa07PASJ:sgrs,lin11ApJ:0501}.

Detailed time-resolved spectroscopy on a large number of bursts
with good statistics have been performed for SGR $1900+14$
\citep{israel08ApJ:1900}, and SGR~J$0501+4516$ \citep{lin11ApJ:0501}.
\citet{israel08ApJ:1900} have studied the 2006 March 29 burst forest
emitted by SGR $1900+14$, when more than 40 bursts were detected by
\swift/BAT, 7 of which were intermediate flares \citep[see also][]{olive04ApJ:1900,
  lenters03ApJ:1900,ibrahim01ApJ:1900}. The authors found that a 2BB
or a COMPT model explains the spectra best. For the 2BB model,
they found average temperatures of $4.8\pm0.3$ and $9.0\pm0.3$~keV,
and radii of $30\pm2$ and $5.7\pm0.5$~km, for the low- and high$-kT$ BB
respectively. The average temperatures and radii for \src\ that we
find are similar within $\sim2\sigma$ uncertainties,
$<$low$-kT>=4.8\pm0.7$~keV, $<$high$-kT>=14\pm2$~keV, and $<R_{\rm
  low-kT}>=20\pm4$~km, $<R_{\rm high-kT}>=3.0\pm1.0$~km.

However, the correlations between the 2BB-model parameters that \citet{
  israel08ApJ:1900} found for SGR $1900+14$ differ slightly from the one
we find here for \src. For instance, \citet{israel08ApJ:1900} found
that the low- and high$-kT$ luminosities contributed almost similarly to
the total energy up to a luminosity of
$\sim3\times10^{40}$~erg~s$^{-1}$, above which the low$-kT$ luminosity
appears to saturate, and the high$-kT$ luminosity increases steadily
to $\sim3\times10^{41}$~erg~s$^{-1}$. We do not see a similar
  effect for \src\ in the lower-right panel of Figure~\ref{ktvskt},
  where the luminosities of the low- and high$-kT$ components are
  positively correlated up to $\sim3\times10^{41}$~erg~s$^{-1}$, an
  order of magnitude higher than the $\sim3\times10^{40}$~erg~s$^{-1}$
  limit (red dotted-dashed line) derived with \citet{israel08ApJ:1900}.

Finally, the $R^2-kT$ correlation that \citet{israel08ApJ:1900} found
is similar to the one we find here but only for their brightest events
with luminosities $>3\times10^{40}$~erg~s$^{-1}$, where the area of
the low$-kT$ decreases with temperature, at a slower pace than the
area of the high$-kT$ component. Below this luminosity, the
relation for both the low- and high$-kT$ areas changes sharply. Although
we see a change in the area$-kT$ correlation with flux for \src\
(Figure~\ref{2BBvsFlux}, lower panels), it is more gradual than the
sharp turnover that \citet{israel08ApJ:1900} reports for SGR $1900+14$.

\citet{lin11ApJ:0501} studied the time-resolved spectra of the
five brightest \fermi/GBM bursts emitted by SGR~J$0501+4516$, fitting
them with a COMPT model and a 2 BB model. The authors found a
clear double correlation between $E_{\rm peak}$ and flux with a break
at a flux value of $\sim10^{-5}$~erg~s$^{-1}$~cm$^{-2}$. Although this
value is consistent with the one we derive for \src\ 
(Figure~\ref{breakVSflux}), this could be a pure coincidence,
considering the difference in distance between the two sources (2 and
5~kpc for SGR~J$0501+4516$ and \src\ respectively\footnote{We note,
  however, that these distances have high uncertainties, rendering
  the comparison of such parameters between the different sources
  problematic.}). Moreover, the slopes of the BPL of the \epeak-flux
relation for SGR~J$0501+4516$ are in agreement, within uncertainties,
to the values we report here for \src\ (Section~\ref{compMod}). For
the 2BB model, \citet{lin11ApJ:0501} found (similar to
\citealt{israel08ApJ:1900} and our results) that the high$-kT$ area
decreases faster than the low$-kT$ area as a function of
temperature. The authors did not find a correlation between $kT$ or
area and flux, although, the quality of the data was not constraining.

Finally, we note that the large amount and superb quality of data that
we have in hand helped reveal new trends between the fit parameters
of the COMPT and, especially, the 2BB model, showing the spectral
evolution during short bursts with unprecedented detail. For the COMPT
model, we established for the first time a strong positive correlation
between the index and flux, above the break flux-value of
$\sim10^{-5}$~erg~cm$^{-2}$~s$^{-1}$; the exact value of the
\epeak-flux relation turnover. Noteworthy new trends for the 2BB model
are, (1) the increase in area with flux, accompanied with a decrease
in temperature, for the high$-kT$ component, (2) the increase in
low$-kT$ area with flux while $kT$ remains constant, until they reach
a common break point in flux where the area starts to saturate and
$kT$ starts increasing, and (3) the smooth change in the slopes of the
BPL fit to the area$-kT$ relation, and disappearing at the lowest
fluxes where a PL is sufficient to explain the area$-kT$ relation,
with a slope of $\sim-4$. We discuss the physical interpretation of
these new results in the context of the magnetar model in the next
section.

\subsection{Radiative mechanism of \src\ bursts}
\label{interpret}

In the following subsections, we interpret our results in the context
of the magnetar model. We would like to note first that the similarity
in the evolution of the 2BB and COMPT model parameters with flux for
all \src\ bursts indicates that they were all triggered, more or less,
by the same mechanism. Moreover, all the correlations that we find
here do not seem to depend on the intrinsic parameters of a given
burst, i.e., independent of the average flux, $T_{90}$, fluence,
etc., implying that the radiative region is developing, and
subsequently cooling, in similar fashion during all bursts.

\subsubsection{Two black body model}

There are several models that describe the triggering mechanism of
magnetar short bursts, most famously, the model of \citet[][TD95
hereafter, see also \citealt{heyl05ApJ:magBurst}]{
  thompson95MNRAS:GF}. TD95 considered the possibility that
departure from magnetostatic equilibrium in the internal magnetic
field of a magnetar, due for instance to Hall drift and ambipolar
diffusion, will cause a build up of crustal stresses. These stresses,
with the presence of the ultra-strong magnetic field, will result in
the cracking of the crust, and the injection of an Alfven wave into
the magnetosphere. Energetic particles are accelerated throughout the
region of the Alfven wave, creating a trapped fireball of photon-pair
plasma in the magnetosphere. Another model for triggering SGR bursts
involves a magnetospheric reconnection caused by magnetic instability
in the magnetosphere \citep{lyutikov03MNRAS:recon,gill10MNRAS:SGRGF}.
This model can lead to the formation of a plasma fireball
trapped inside magnetospheric flux lines \citep{gill10MNRAS:SGRGF}, and
also to thermal emission from heating of the magnetar surface. We note
that the crust cracking model can lead to significant magnetospheric
reconnection, since the external magnetic field will feel any
instability in the internal one.

Regardless of the triggering mechanism, the trapped fireball formed
during SGR bursts is hot, $T\gtrsim50~$keV, with an extremely high
density (TD95). Hence, the radiative mechanism of short SGR bursts is
the cooling of a hot, optically thick plasma, confined by an
ultra-strong magnetic field (B$>$B$_{\rm QED}$). In this regime, the
dominant energy exchange process is Compton scattering, and due to the
high optical depth in the fireball, photons will most likely be
thermalized. The diffusion time-scale across the fireball at a 
distance $R$ from the NS surface is much larger ($\sim10^4$~s, TD95)
than the typical SGR burst time-scale ($\sim0.2$~s) due to the
enormous optical depth through the pair plasma. Hence, the fireball
likely loses energy as its cool surface layer propagates inwards
(TD95). A gradient in the magnetic field is expected throughout the
fireball due to its large coronal volume ($B\propto R^{-(2+p)}$,
$0<p<1$ depending on the twist of the external dipole field, and $R$
is the distance from the NS surface,
\citealt{thompson02ApJ:magnetars}), causing a gradient in the
effective temperature of the fireball at different heights above the
NS surface (TD95). The true spectrum would then be a distorted
multi-color BB.

Alternatively, \citet[][see also \citealt{kumar10ApJ:0501}]{
  israel08ApJ:1900} attributed the 2BB model, which they used to fit
the SGR~$1900+14$ bursts, to the effect of the strong magnetic field
($B>B_{\rm QED}$) on the scattering cross-sections of the two
polarization states of the photons \citep{herold79:qed,
  meszaros80ApJ:radTrans}. In super-strong magnetic fields, E-mode
photons (photons with electric vector perpendicular to B) have a much
reduced cross-section and can potentially diffuse out from deep inside
the magnetosphere very close to the NS surface, equivalent to the
high$-kT$ component. O-mode photons, on the other hand, have a much
higher cross-section and diffuse out from further out in the
magnetosphere at large radii, correpsonding to the low$-kT$ component
(TD95, \citealt{lyubarsky02MNRAS:bursts}). However, \citet{
  vanputten13MNRAS:magAtmos}, modeling hydrostatic atmospheres in
super-strong magnetic fields, showed that the E- and O-mode
photospheres are very close in both temperature ($\Delta kT_{\rm
  E-O}\sim1$~keV) and location, in sharp contrast to our findings (and
those of \citealt{israel08ApJ:1900}) between the low- and high$-kT$
components. We note that the true picture is even more complicated due
to the fact that the evolving fireball during a given burst will
sample a large range of field strengths and orientations, thereby
mixing the radiative transfer elements of Compton opacity.

One way of overcoming the complications mentioned above, is if the 2BB
components represent two physically, and geographically distinct
emitting regions, each with their own E-mode and O-mode
photospheres. According to the distribution of areas and temperatures
of our 2BB modeling, one would expect a hot one located close to the
NS surface (or a hot spot on the surface), and a cooler one located
further out in the magnetosphere. The large spatial separation between
the two emitting areas ($\sim$2 orders of magnitude in $R^2$) will
most likely induce a profound difference in magnetic field morphology
between these two zones. Nonetheless, the strong positive correlations
that we find between the areas and fluxes of these two regions
indicate that they are very strongly connected, and emitting at
similar rates.

How do these two emitting regions evolve during a burst? For the
high$-kT$ emitting region, the decrease in temperature (although
small) and increase in area with flux (Figure~\ref{2BBvsFlux},
upper-panels, which is also true within each burst) is a clear
indication of an adiabatically expanding emitting region, in line with
the fireball model. The area of this emitting region ranges from a
small hot spot on the NS surface ($\sim0.2$~km) up to a radius of
$\sim5$~km. This radius is consistent with the size of the effective
radiative zone of the footpoints of the plasma fireball according to
TD95 (considered to be the E-mode photosphere), which is about $R_{\rm
  NS}/2$. Also, the temperature of this emitting region varies
slightly, ranging on average from 10~keV to 15~keV
(Figures~\ref{ktdist} and \ref{2BBvsFlux}), although fluxes vary by
more than 3 orders of magnitude. This is also in line with the cooling
fireball model, which predicts a slight spectral variation of the
emergent spectrum from the emitting region (TD95).

The evolution of the low$-kT$ component with flux is not straightforward
to interpret. The temperature of this emitting region appears constant
at very low fluxes while the area increases at a similar rate as the
high$-kT$ area (i.e., with the same slope of $\sim1$,
Figure~\ref{2BBvsFlux}, upper-panels). At
$F\approx10^{-5.5}$~erg~s$^{-1}$~cm$^{-2}$, the area starts to
saturate while the temperature starts increasing. Since both BB
components emit radiation at the same rate (Figure~\ref{ktvskt},
lower-left panel), an increase in temperature for the low$-kT$ component 
with increasing flux, while its area starts saturating, is inevitable in
order to maintain the same energy density as the high$-kT$ component.

At the highest fluxes, the radius of the low$-kT$ emitting area seems to
reach a maximum of $R_{\rm max}\sim30$~km (see Section~\ref{2BBmod},
and upper-left panel of Figure~\ref{2BBvsFlux}). This radius is much
bigger than the NS radius (assuming surface emission), even if we
include the effect of gravitational lensing, $R_{\rm
  app}=R(1-2GM/Rc^2)^{-1/2}$, where $R_{\rm app}$ is the apparent
radius of the NS at infinity, $R$ and $M$ are its radius and mass
\citep{pechenick83ApJ:lens, psaltis00ApJ:lens, ozel13RPPh:reviewNS}.
In fact, considering even the largest and most massive neutron stars
allowed by theoretical equation of states \citep{lattimer01ApJ:EOS},
we have $R_{\rm app}\approx20$~km (Figure~3 in \citealt{
  lattimer12ARNPS:EOS}), much lower than the saturation radius of the
low$-kT$ emitting region. On the other hand, in the context of the
fireball model, where the surface area $A_{\rm max}=\pi R_{\rm max}^2$
is a closed magnetospheric magnetic field bundle that anchors the
emitting plasma, the saturation at such a level is expected since, for
obvious geometrical reasons, it is difficult for the emitting region
to be significantly larger than the NS area.

In the crust-cracking model for triggering SGR bursts (TD95), this 
saturation radius, $R_{\rm max}\sim30$~km, of the high$-kT$ component
has very important consequences. The frequency of the Alfven wave
responsible for the creation of the photon-pair plasma fireball is,
$\nu_{\rm max}=c/R_{\rm max}\approx10^4$~Hz\footnote{We note that the
  wave Alfven speed is relativistic, since the magnetic energy density
  at $R_{\rm max}$ is greater than the corresponding maximum energy
  density of the burst. $B_{\rm R\_max}^2/8\pi\approx4\times10^{20}$
  (considering $B_{\rm R\_max}=10^{11}$~G, see above), while $E_{\rm
    max}/4.2R_{\rm max}^3\approx10^{19}$~erg~cm$^{-3}$, with $E_{\rm
    max}=1.2\times10^{39}$~erg.} (which is consistent with a
characteristic seismic mode frequency for neutron star crusts;
\citealt{blaes89ApJ:NSquakes}). This harmonic excitation of the
magnetosphere, like any other, has a minimum characteristic excitation
radius, $R_\nu$ (TD95). This excitation radius cannot exceed the
radius of the magnetic loop, $R_{\rm Loop}$, trapping the plasma
fireball, $R_\nu<R_{\rm Loop}$, as this will result in an inefficient
excitation (TD95). 

The maximum effective area of the low$-kT$ component, $\pi R_{\rm
  max}^2$, could be assumed to be a fraction, $l_5/R_*$, of the
expected projected area of the relevant magnetic field loop, $\pi
R_{\rm Loop}^2$, where $l=l_5\times10^5$~cm is the length-scale of the
surface crack (TD95), and $R_*$ is the NS radius (assumed to be
10~km). This implies that

\begin{equation}
R_{\rm Loop} \sim (R_*/l_5)^{1/2} R_{\rm max} \sim 100~{\rm km,}
\end{equation}

\noindent  assuming $R_{\rm max}=30$~km. This $l_5/R_*$ scaling comes
from the fact that the derived radius, $R_{\rm max}$, of the low$-kT$
effective area spans magnetic field lines anchored over a large angle
of order unity on the stellar surface, whereas in the case of a
magnetic loop with small area footpoints (corresponding to the hot BB
effective area), a smaller angle of order $\sim l_5/R_*$ is
expected. Finally, according to the star-quake model, the minimum
excitation radius of the Alfven wave is (TD95),

\begin{equation}
R_\nu \sim 10~B_{15}^{-2} \left(\frac{\theta_{\rm
      max}}{10^{-3}}\right) \left(\frac{V_\mu}{1.4\times10^8 {\rm~cm~s^{-1}}}\right)^{-1}~l_5~{\rm km},
\end{equation}

\noindent  where $B=B_{15}\times10^{15}$~G is the core magnetic field
strength, $\theta_{\rm max}\sim10^{-3}-2\times10^{-1}$ is the yield
strain of the crust \citep{horowitz09PhRvL:magStrain}, and
$V_\mu\sim(0.4-1.4)\times10^8$~cm~s$^{-1}$ is the shear wave velocity
\citep{steiner09PhRvL:NSEOS,douchin01AA:NSEOS}. As mentioned above,
since $R_\nu<R_{\rm Loop}\sim100$~km, this yields a crustal internal
magnetic field, 

\begin{equation}
\begin{split}
B & \gtrsim 3.2\times10^{14} \left(\frac{R_{\rm Loop}}{100 {\rm~km}}\right)^{-1/2} \left(\frac{\theta_{\rm max}}{10^{-3}}\right)^{1/2}\\
   & \left(\frac{V_\mu}{1.4\times10^8 {\rm~cm~s^{-1}}}\right)^{-1/2} l_5 ^{1/2}~{\rm G}.
\end{split}
\end{equation}

We note that we used the limits on $V_\mu$ and $\theta_{\rm max}$ that
yield the stringent lower limit on the crustal internal magnetic field
$B$. A more reasonable values for $V_\mu$ ($=0.7\times10^8$~cm~s$^{-1}$,
\citealt{steiner09PhRvL:NSEOS}), and $\theta_{\rm max}$ ($=0.1$, \citealt{
  horowitz09PhRvL:magStrain}), and substituting $R_{\rm Loop}$ with
$R_{\rm max}$ results in,

\begin{equation}
\begin{split}
B & \gtrsim 4.5\times10^{15} \left(\frac{R_{\rm max}}{30 {\rm~km}}\right)^{-1/2} \left(\frac{\theta_{\rm max}}{0.1}\right)^{1/2}\\
   & \left(\frac{V_\mu}{0.8\times10^8 {\rm~cm~s^{-1}}}\right)^{-1/2} l_5 ^{1/2}~{\rm G}.
\end{split}
\end{equation}

We could also derive an upper-limit on the crustal internal magnetic
field considering that the maximum observed energy, $E_{\rm
  max}=L_{\rm max}\times4_{\rm ms}\sim1.2\times10^{39}$~erg, where
$4_{\rm ms}$ is the bin time-interval, does not exceed the maximum
magnetic energy possibly released in an area $l_5^2$ on the crust,
which is $E_{\rm max}\sim4\times10^{40}~l_5^2~B_{15}^{-2}~(\theta_{\rm
  max}/10^{-3})^2$~erg (TD95). Taking $\theta_{\rm max}=0.1$, the above
implies,

\begin{equation}
B \lesssim 5.8\times10^{17} \left(\frac{\theta_{\rm
      max}}{0.1}\right) l_5~{\rm G}.
\end{equation} 

Another interesting finding in our analysis is the $R^2-kT$
relation, and its dependence on flux, which is once again a
complicated result to interpret qualitatively (Figure~\ref{2BBvsFlux},
lower-panels). For instance, at the lowest fluxes
($F<10^{-5.5}$~erg~s$^{-1}$~cm$^{-2}$) we find that $R^2\propto
kT^{-3.8\pm0.4}$, i.e., following lines of constant luminosities 
($F\propto R^2kT^{4}$). The relation breaks from the above with
increasing flux. This indicates that at the lowest fluxes, photons
emitted by both components follow a perfect Planckian distribution,
and starts deviating from it with increasing flux. One possible way to
achieve a purely thermal spectrum is if photon splitting
\citep{adler70PhRvL:phSplit,baring95ApJ:phSplit,
  baring01ApJ:phSplit,usov02ApJ:phSplit,chistyakov12PhRvD:phSplit} is
the dominant source of photon number changing (TD95). At the lowest
fluxes, the size of the high$-kT$ component is extremely small (some
tens of meters), and very likely most of the emitted radiation is in
the form of E-mode photons. At such small distance from the NS, the
magnetic field is super-critical, and due to the high temperature of
the high$-kT$ component at these lowest fluxes, photon splitting is
very efficient (low$-kT\sim14$~keV$>kT_{\rm sp}\sim11$~keV, where
$kT_{\rm sp}$ is the BB temperature at which the photon splitting rate
is sufficient to maintain LTE, TD95), leading to complete
thermalization of the high$-kT$ component. With increasing flux, the
high$-kT$ temperature starts dropping and the spectrum starts deviating
from a pure Planckian, perhaps due to other radiative processes coming
into play, e.g., double Compton scattering ($e+\gamma\rightarrow
e+\gamma+\gamma$) and Bremsstrahlung. For the low$-kT$ component,
the picture is not clear, mostly due to the variation in field
morphology and opacity for both scattering and photon splitting at
high altitudes. However, since these two components appear to be
strongly coupled (Figure~\ref{ktvskt}), we speculate that a deviation
in the high$-kT$ photon distribution from a pure Planckian model would
lead to the deviation in the low$-kT$ component as well.

\subsubsection{Comptonized model}

The COMPT model that we use here in our
time-resolved spectroscopy is meant to mimic classical problems of
unsaturated Comptonization (see \citealt{lin11ApJ:0501} for more
details). In this model, photons scatter repeatedly inside a corona
of hot electrons, until they reach the plasma temperature, $E\sim
kT_{\rm e}$. Further heating is impossible and a spectral turnover
(\epeak) emerges. The power-law index below the turnover ($\lambda$)
depends on the mean energy gain per collision and the probability of
photon loss from the bubble \citep{rybicki79rpa:radProcess}. Hence,
the index $\lambda$ depends only on the magnetic Compton$-y$ parameter
($y_{\rm B}$), $\lambda=1/2-\sqrt{9/4+4/y_{\rm B}}$, with $y_{\rm
  B}=4kT_{\rm e}/(m_{\rm e}c^2){\rm max}\{\tau_{\rm B},\tau_{\rm
  B}^2\}$. Here $\tau_{\rm B}$ is the effective optical depth inside
the scattering medium, modified by the strong magnetic field, and
${\rm max}\{\tau_{\rm B},\tau_{\rm B}^2\}$ is the mean number of  
scatterings per photon. 

We note that in magnetars, such a corona of hot electrons could
develop in the inner magnetosphere, caused by the twisting of the
external magnetic field lines \citep{beloborodov07ApJ:magCorona,
  thompson02ApJ:magnetars, thompson05ApJ:magCorona,
  nobili08MNRAS:XraySpec}. The magnetic reconnection model of
\citet{lyutikov03MNRAS:recon} for triggering SGR bursts fits this
picture well, with any reconnection event being accompanied with
magnetic field line twists.

The correlation that we find between the index $\lambda$ and flux is
interesting in the context of the COMPT model
(Figure~\ref{indexVSflux}). At low fluxes,
$F\lesssim10^{-5}$~erg~s$^{-1}$~cm$^{-2}$, $\lambda$ is distributed
around $\sim-1$, i.e., the flattest possible spectra. Such spectra are
achieved through repeated Compton upscattering with $y_{\rm
  B}>>1$. However, with increasing fluxes, $\lambda$ starts increasing
to reach $\sim1$ at the highest fluxes. Compton upscattering of soft
photons has difficulty generating such spectra. These high values for
$y_{\rm B}$ suggest high opacity, and strong thermalization might be
taking place inside the corona. Moreover, the \epeak-flux relation
enforces this conclusion. The anti-correlation that we see
between these two parameters below the break at
$F\approx10^{-5}$~erg~s$^{-1}$~cm$^{-2}$ (Figure~\ref{breakVSflux},
which roughly corresponds to the break we see in the $R^2-$flux
and $kT-$flux relations for the low$-kT$ component) could be explained
if there were coronal radius expansion (exactly the case of the
high$-kT$ BB component, see above). On the other hand, the switch in
this relation above the break and the hardening of the spectra with
flux, indicates a saturation radius of the coronal volume, again
corresponding to the saturation radius that we see for the low$-kT$ 
component. In the light of the above results, there is no reason not
to prefer a true thermal spectrum at all fluxes for \src\ bursts,
i.e., throughout the evolution of a given burst. Similar  results were
derived by \citet{lin12ApJ:1550} when studying the broadband,
1$-$200~keV, time-integrated spectra of \src\ bursts.

\section{Concluding remarks}
\label{conclusion}

In this paper, we have shown the importance of high time-resolution in
the study of the spectral evolution of SGR bursts. We were able to put
to test the emission from a hot, optically thick, fireball of
photon-pair plasma, predicted to form during SGR short bursts,
according to a number of models for triggering magnetar flares (TD95,
\citealt{gill10MNRAS:SGRGF}, \citealt{heyl05ApJ:magBurst}). We were
able to follow the evolution of the emitting plasma in an
unprecedented detail. Nonetheless, the very detailed observational
picture presented in this paper motivates a more in depth modeling of
the development and evolution of scattering/splitting cascades
involving photon-pair fireballs in magnetar magnetospheres, taking
into account the many physical and geometrical effects, to better
understand the emission mechanism taking place during magnetar
bursts.

\section*{Acknowledgments} This publication is part of the
GBM/Magnetar Key Project (NASA grant NNH07ZDA001-GLAST; PI:
C. Kouveliotou). ALW acknowledges support from an NWO Vidi grant.
AJvdH and RAMJW acknowledge support from the European Research Council
via Advanced Grant no. 247295. We thank the referee for constructive
comments that helped improve the quality of the manuscript.

\end{document}